# Improved absolute abundance estimates from spatial count data with simulation and microfossil case studies

Chris Mays[1]*, Marcos Amores[1], Anthony Mays[2]

[1]School of Biological, Earth and Environmental Sciences, Environmental Research Institute, University College Cork, Distillery Fields, Cork T23 N73K, Ireland

[2]School of Computer and Mathematical Sciences, University of Adelaide, Adelaide 5005, Australia

*Corresponding author

E-mail: cmays@ucc.ie

Short title: New absolute abundance spatial count method

## Abstract

Many fundamental parameters of biological systems—such as productivity, population sizes and biomass—are most effectively expressed in absolute terms. In contrast to proportional data (e.g., percentages), absolute values provide standardised metrics on the functioning of biological entities (e.g., organisms, species, ecosystems). These are particularly valuable when comparing assemblages across time and space. Since it is almost always impractical to count entire populations, estimates of population abundances require a sampling method that is both accurate and precise. Such absolute abundance estimates typically entail more 'sampling effort' (= data collection time herein) than proportional data. Here, we refined a method of absolute abundance estimates—the 'exotic marker technique'—by producing a variant that is more efficient (increased precision, reduced effort), without losing accuracy.

This new method, the 'field-of-view subsampling method' (FOVS method), estimates absolute abundances across spatial data sets. The FOVS method is based on area subsampling, from which large samples can be quickly extrapolated, but with an additional source of error (from variations in specimens per field of view). Two case studies of the exotic marker technique were employed: 1, computer simulations; and 2, an observational 'real world' data set of terrestrial organic microfossils from the Permian- and Triassic-aged rock strata of southeastern Australia. To serve as exotic markers, microfossil samples were spiked with doses of *Lycopodium* spores of known quantity and variance. To compare the FOVS method against the traditional method (the 'linear method' herein), three output parameters were measured: 1, concentration (= specimens/gram of sediment, in the microfossil case study); 2, precision and 3, data collection effort. Our findings demonstrate how vital the ratio between targets to markers in an assemblage is for achieving high-precision estimates. In almost all cases, the FOVS method delivers higher precisions than the linear method, with equivalent effort. In contrast, our computer simulations revealed that high target-to-marker ratios significantly impact not only the precision, but also the accuracy of the precision estimates, of the linear method. The linear method had superior efficiency only for assemblages with very low specimen densities and/or near-equivalent target-to-marker ratios. Since we predict that these conditions are typically very rare, we recommend the new FOVS method in almost every 'real world' case.

Ostensibly, these count methods have been developed for microfossil-based data sets; however, they are amenable to many area-based count data where markers of known quantity can be introduced into a population. We provide guidelines and a user-friendly interface to aid in implementing this new count technique, as well the simulation model codes for readers to inform their own experimental designs.

## Introduction

Absolute abundances are key parameters across the biological sciences, including conservation (e.g., species population sizes [1]), biomedicine (e.g., cell counts [2]), agriculture (e.g., herd sizes [3]) and ecology (e.g., biomass [4]; primary productivity [5]). A key strength of absolute abundances is that they are divorced from the issues inherent to proportional (or relative, e.g., percentage) abundance data, such as compositional effects [6–8]. A common compositional effect occurs when a relative increase in one component necessarily results in a decrease of all other components, regardless of whether there is a genetic link between them; hence, the analyses of such data can lead to spurious correlations [9].

When applied to past biological systems, fossil abundance data are used to infer organism spatiotemporal distributions, which can inform myriad environmental, ecological and evolutionary trends. Relative data can clearly indicate population changes in Earth's past [10, 11], but are most validly applied to limited spatiotemporal contexts (i.e., similar areas and time ranges). However, by providing standardised benchmarks, absolute abundances enable valid comparisons between vastly different spatiotemporal contexts, while facilitating additional inferences beyond population shifts. One common absolute abundance metric for past ecosystems, fossil concentration (= number of fossils per unit mass; "$c$" herein), quantifies the specific biological contributors to a sedimentary rock [12]. Such absolute fossil abundances are particularly important during past, rapid environmental changes. This is owing to the extreme abundance fluctuations of various fossil groups (e.g., algae or 'acritarchs', [13–15]; fungi, [16, 17]; fern spores, [18, 19]; charcoal, [20, 21]). However, apparent proliferation events or abundance 'spikes' based on relative data might simply reflect the extirpation of other fossil groups. Absolute metrics can indicate whether group abundances are correlated or independent, which is a crucial step in drawing conclusions about causation regarding spatiotemporal trends in Earth's history.

The 'exotic marker technique' is a method for measuring absolute abundances, and performed by simultaneously counting samples from populations of: 1, target specimens; and 2, 'exotic markers' (of known abundance). Crucially, the markers and targets must behave similarly, to result in homogeneous mixing of the two populations. As highlighted by Maher [22], this technique has long been utilised in disparate fields, such as agriculture (e.g., counting black sheep among predominantly white flocks) and ecology (e.g., releasing tagged animals into the wild and later counting the ratio of tagged vs non-tagged individuals). First applied to organic microfossils by Benninghoff [23], the exotic marker technique has undergone a series of iterations with different markers [24, 25], including distinctive angiosperm pollen (e.g., *Ailanthus altissima* (Mill.) Swingle: [26]; *Alnus incana* (Du Roi) R.T. Clausen: [27, 28]; *Eucalyptus globulus* Labill.: [29–32]; *Nyssa sylvatica* Marshall: [33]), microscopic plastic beads (e.g., [34–37]) or ceramic microspheres [38]. However, the most widely utilised exotic markers are spores of *Lycopodium clavatum* L. [25, 39–42], which have been primarily applied to Quaternary sediment samples (e.g., [43–46]) but increasingly to older assemblages (e.g., [47–49]). Similar approaches have been successfully applied to other organism groups (e.g., diatoms; [50–52]). Whichever exotic marker is used, ideally they should be optically distinct from the indigenous specimens, whilst having similar behavioural properties (e.g., the same hydrodynamic and chemical properties as the microfossil targets) to avoid overrepresentation of either group [53].

This study aims to statistically assess and refine the 'exotic marker technique', a commonly utilised measure of absolute abundances. Two parallel data sets—computer-based simulation and organic microfossil assemblages—will be collected to test a new variant of this technique, and compare it to the traditional method (the 'linear method' herein). The new variant (the 'field-of-view subsampling method', or FOVS method) has the potential to provide greater efficiency (improved precision and/or reduced data collection effort) and increase the versatility of legacy samples or previously collected spatial data. Further, we aim to maximise the utility of the count methods by providing: 1, precise criteria for choosing the optimal method for each sample; 2, the source code for the simulated data sets; and 3, an accessible, online interface for estimating key parameters and employing both methods.

## Materials and methods

## Specimen absolute abundances

The exotic marker technique measures the absolute abundances of the specimens of interest, or 'targets'. A common absolute abundance is specimen concentration ($c$), measured per unit size (e.g., mass, area or volume). For concentration estimates, one must: 1, measure the original sample size; 2, add a quantity of exotic markers with known mean and uncertainty (e.g., standard deviation); and 3, compare the counted abundances of both targets and exotic markers from a sample. Another way to express this: since we know the number of introduced marker specimens per unit size (i.e., exotic marker concentration), we can infer the concentration of indigenous specimens by measuring the ratio of target to marker specimens.

The general formula for these concentrations ($c$) follows Benninghoff [23], with terms updated from [22]:

$$c = \frac{x \times N_1 \times \overline{Y}_1}{n \times \overline{V}} \tag{1}$$

where $x$ = the total target (e.g., microfossil) specimens counted, $N_1$ = the number of doses (e.g., tablets) of exotic markers added to the sample, $\overline{Y}_1$ = mean number of exotic markers per dose, $n$ = exotic marker (e.g., *Lycopodium* spore) specimens counted, and $\bar{V}$ = total size (mass, area or volume) of sample. Unless specified, $c$ indicates concentrations derived using the originally proposed concentration method, dubbed the 'linear method' herein (see 'linear method: operation' below). (The term for linear method concentrations may also take the form $c_L$ to differentiate it from $c_F$, the latter of which is the concentration value from the newly proposed 'FOVS method').

Significance tests were conducted with PAST v. 4.13 [54]. A glossary of terms, abbreviations and initialisms used in this study is provided in S2 Table. Specifics of computer simulation parameters are outlined in the 'case study: Computer simulation' section. Specifics of microfossil sample processing, and the application of microfossil concentrations techniques for estimating terrestrial sedimentary organic matter are outlined in the 'case study: Permian–Triassic of eastern Australia' section. Additional parameter corrections are included in S1 Text.

## Counting techniques

In this study, we tested two distinct count methods for specimen concentrations: 1, the linear method; and 2, the field-of-view subsampling (FOVS) method. The following sections outline the rationale and practical aspects of each, and introduce their key parameters. However, these sections assume that the choice of method has already been made. The method determination process is summarised in steps 1 and 2 of Fig 1, but is discussed in detail later (section 'choosing the superior count method') since this requires knowledge of the key sample and count parameters discussed in the immediately following sections.

### Linear method: operation

To estimate specimen concentrations in a population, the 'linear method' has been commonly used (Eqn 1), whereby concentrations of thousands or even millions of specimens (per unit mass or volume) can be estimated from only a few hundred. This method involves the identification and counting of individual specimens, and their assignment to two (or more) specimen categories: markers ($n$) and targets ($x$ or, for multiple target categories, $x_1, x_2, \dots x_k$).

This can be considered a linear procedure, whereby specimen counts (both $x$ and $n$) are collected simultaneously and sequentially. In the simplest case, this is conducted until a pre-determined minimum value of $x$ is reached (although one could also use pre-determined $n$ values). Stopping when a predetermined number of target specimens is reached defines a random “window” of a sample area (e.g., microscope slide) within which we now have a random number of markers. In the microfossil case study, the target population is the total number of microfossils in a given sediment, and the marker population is the total number of exotic markers (e.g., *Lycopodium* spores) introduced to the sample. Since all target specimens in a sample are examined individually and assigned to a count category, the linear method tends to provide accurate identifications; however, the process of individual specimen identification can be quite time-consuming, particularly with higher numbers of target categories.

The highest precisions for concentration estimates are achieved when the target specimens and exotic markers are equal in the population (i.e., $x = n$; Regal & Cushing [55]). We recover this result as a by-product of mathematical analyses comparing the efficiency of the two methods (see S1 Text). However, for combined ($x + n$) specimen counts of several hundred (e.g., ≥500), a ‘target-to-marker ratio’ (or $\hat{u}$, following the terminology of [22]) close to 2:1 provides the most efficiency by striking the optimum balance between precision and data collection effort [22]. For this target-to-marker ratio ($\hat{u} \approx 2$), count sizes greater than 500 provide sharply diminishing returns with increased effort (translating to increased data collection time). Similarly, Price et al. [42] recommended that the target-to-marker ratio should be <5:1 (or $\hat{u} < 5$). In cases where the target and marker ratios are near-equivalent (for single target counts), precise concentrations can typically be achieved with target counts of less than 300 [40].

(Note: unless stated otherwise, we assume that the targets are more common than the exotic markers; hence, we utilise the term 'target-to-marker ratio' in place of 'common-to-rare ratio'. If, however, the markers are more common than the targets, see the relevant section in the S1 Text.)

## Linear method: precision estimates

To gauge the precision of concentration estimates using the linear method, concentration total error values ($\sigma_L$) were calculated with the formula (from [25], updated with terms defined by [22]):

$$\sigma_L = 100\sqrt{\left(\frac{s_{1P}}{\sqrt{N_1}}\right)^2 + \left(\frac{\sqrt{x}}{x}\right)^2 + \left(\frac{\sqrt{n}}{n}\right)^2} \qquad (2)$$

where $\sigma_L$ is the total standard error of the linear method concentration estimate (in %). In this formula, the variables $x$, $n$ and $N_1$ are as in Eqn 1, while $s_{1P}$ is the proportional sample standard deviation of the number of exotic markers per dose (see S2 Table).

The standard deviations ($\sqrt{x}$ and $\sqrt{n}$) come from the underlying assumption that all specimens (targets and markers) are independently distributed according to a Poisson point process (i.e., they are distributed uniformly at random over the study area and, hence, their distributions are independent of each other). These assumptions provide a valid statistical framework for many data in the biological [56] and physical [57] sciences. The standard deviations are then divided by the respective number of counted specimens to yield proportional standard deviations. Note that since the number of targets ($x$) (or, in some cases, the number of markers [$n$]) is pre-determined in the linear method, the inclusion of error terms for both may seem counterintuitive. However, the error term coming from the fixed quantity can be understood as the contribution to the error from the variable size of the linear method window determined by the random locations of the counted specimens.

Since the specimens are counted sequentially, neither density nor homogeneity (defined here as the variance of the specimen density across the study area) will impact the precision of this method. However, these factors will certainly affect the 'data collection effort', which is reflected by the time spent during counting (e.g., at the microscope, sparsely populated slides will take a longer time to achieve a given count size). We have not provided a comprehensive discussion of the effects of inhomogeneity, since we do not yet have quantitative data to test this; however, this would serve as the basis for future investigation.

Precision of linear method concentration was also estimated by calculating confidence intervals for $c_L$. For these, the parametric approach by Maher (1981, p. 179) was implemented, the relevant parameters and formulae for which are summarised in S2 Table. A minor change to the confidence interval calculation was made to adapt it to sample mass, not volume, following [40]. Maher's [22] calculations encompass the uncertainty in size (in units of volume or mass) and number of samples. In the microfossil case study below, the masses were considered true values, and standard errors were set to 0.1 g (equivalent to the precision limits of the mass measurement).

## Field-of-view subsampling (FOVS) method: operation

The FOVS method is a count data collection approach that we have designed as an alternative, or complement, to the linear method. It extrapolates specimen abundances from preset surface areas. This area-based sampling is distinct from the linear method above, where both target ($x$) and marker ($n$) specimens are counted in a continuous sequence. Instead of linear, stepwise counts of all individual specimens, the FOVS method obtains large abundance data sets by extrapolating abundances from surface area subsamples.

Analogous area-based subsampling techniques have been applied widely across disparate sciences to infer population statistics [58], but—to our knowledge—never before to organic microfossil assemblages (cf. [52]). For example, quadrat sampling is a standard method used in field ecology, whereby the specimens within a set of representative areas are counted, from which populations can be extrapolated across a broader study area [59, 60]. Within the context of microfossil analyses, we propose that the microscope field of view (FOV) can be utilised as the unit of surface area, in place of the quadrat. For valid results, the subsample areas or 'fields of view' should be consistent for all counts of a given assemblage.

The essential difference between standard quadrat methods and the FOVS method proposed herein is the addition of exotic markers of known quantity (Eqn 1). From these, the observed proportion of targets to markers provide an accurate assessment of total target concentrations within a population. This combination of concentration estimation and surface area subsampling holds the potential for precise absolute abundances with minimal sampling effort.

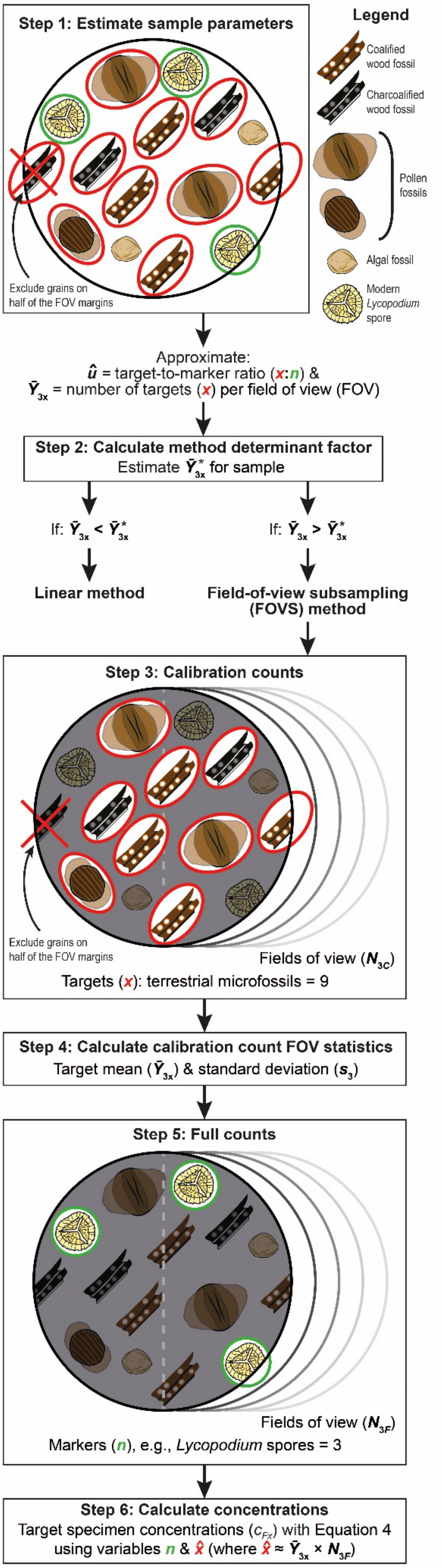

Step 1: Estimate sample parameters
Legend
Coalified wood fossil
Charcoalified wood fossil
Pollen fossils
Algal fossil
Modern Lycopodium spore
Exclude grains on half of the FOV margins
Approximate:
$\hat{u}$ = target-to-marker ratio ($x$:$n$) &
$\bar{Y}_{3x}$ = number of targets ($x$) per field of view (FOV)
Step 2: Calculate method determinant factor
Estimate $\bar{Y}^*_{3x}$ for sample
If: $\bar{Y}_{3x} < \bar{Y}^*_{3x}$
If: $\bar{Y}_{3x} > \bar{Y}^*_{3x}$
Linear method
Field-of-view subsampling (FOVS) method
Step 3: Calibration counts
Exclude grains on half of the FOV margins
Fields of view ($N_{3C}$)
Targets ($x$): terrestrial microfossils = 9
Step 4: Calculate calibration count FOV statistics
Target mean ($\bar{Y}_{3x}$) & standard deviation ($s_3$)
Step 5: Full counts
Fields of view ($N_{3F}$)
Markers ($n$), e.g., Lycopodium spores = 3
Step 6: Calculate concentrations
Target specimen concentrations ($c_{Fx}$) with Equation 4
using variables $n$ & $\hat{x}$ (where $\hat{x} \approx \bar{Y}_{3x} \times N_{3F}$)

**Fig 1. A flowchart of two count methods for estimating absolute abundances using exotic markers, with an organic microfossil case study.** The flowchart includes a step-by-step procedure of the field-of-view subsampling (FOVS) method. To calculate $\overline{Y}^*_{3x}$ for determining the most efficient method for each sample, use Eqn 20 when $x > n$; if $x < n$, we recommend calculating $\overline{Y}^*_{3n}$ instead with S12 Eqn (see S1 Text for more details). All calculations can be made with the user-friendly data collection interface (https://github.com/palaeomays/FOVS_vs_linear_methods.git), including the optimal numbers of calibration-count and full-count fields of view ($N^*_{3C}$ and $N^*_{3F}$, respectively). Note: an additional input parameter for Eqn 20 is the researcher-specific time parameter $\omega$ (= ratio of time taken for field-of-view transitions vs individual specimen counts). FOV = field of view.

The FOVS method involves the following sequence of steps, which are summarised in Fig 1; see S2 Table for equations and terms. (Note: The following procedure assumes that the FOVS method has been determined as the most appropriate for the given population [Fig 1, steps 1, 2]. For more details on the method determination process, see discussion in 'choosing the superior count method').

A. Examine the assemblage to determine whether the targets or markers are more common. (For the following example, the target specimens, $x$, are assumed to be more common. If markers, $n$, are more common than $x$, see the S1 Text.)
B. Conduct a series of 'calibration counts' for the target specimen type (Fig 1, step 3). These consist of counts of all target specimens in a subsample of fields of view (where $N_{3C}$ is the total number of calibration-count fields of view). To avoid overrepresenting the number of target specimens per unit area, exclude half of the specimens on the field-of-view margins (e.g., on one side of the field of view; cf. [61]).

C. Calculate the mean ($\overline{Y}_3$) and standard deviation ($s_3$) of the target specimens per field of view from the calibration counts (Fig 1, step 4).

D. Conduct a 'full count' of new fields of view (Fig 1, step 5), whereby two values are collected simultaneously: 1, the number of full-count fields of view ($N_{3F}$); and 2, the number of rare specimens (typically, the exotic markers, $n$). To ensure that the calibration counts are representative, the regions chosen for both the full and calibration-count fields of view should be located close to each other on the microscope slide. In this way, the specimen density and heterogeneity will be approximately equivalent (Fig 2A).

E. Organic microfossil concentration can then be calculated (Fig 1, step 6). Assuming the target specimens were the common specimens during the calibration counts, then an approximate value for $x$ can be calculated by multiplying the mean number of targets per field of view from the calibration counts ($\overline{Y}_{3x}$) by the counted number of fields of view from the full counts ($N_{3F}$). This extrapolated value of $x$ for the full count is denoted $\hat{x}$, i.e.,

$$\hat{x} = \overline{Y}_{3x} \times N_{3F}. \tag{3}$$

The subscript $x$ in the term $\overline{Y}_{3x}$ indicates that the target specimens are the most common specimen type, and subjects of the calibration counts. The total number of exotic markers from the full counts ($n$) can then be inserted directly into a modified version of Eqn 1:

$$c_{Fx} = \frac{\hat{x} \times N_1 \times \overline{Y}_1}{n \times \overline{V}}, \tag{4}$$

where we denote the estimated organic microfossil concentration from the FOVS method by $c_{Fx}$ to distinguish it from the linear method concentration estimate ($c_L$; see Eqn 1). (The subscript "$x$" for $c_{Fx}$ assumes that the target specimens are the foci of the calibration counts; if this is not the case, see the S1 Text).

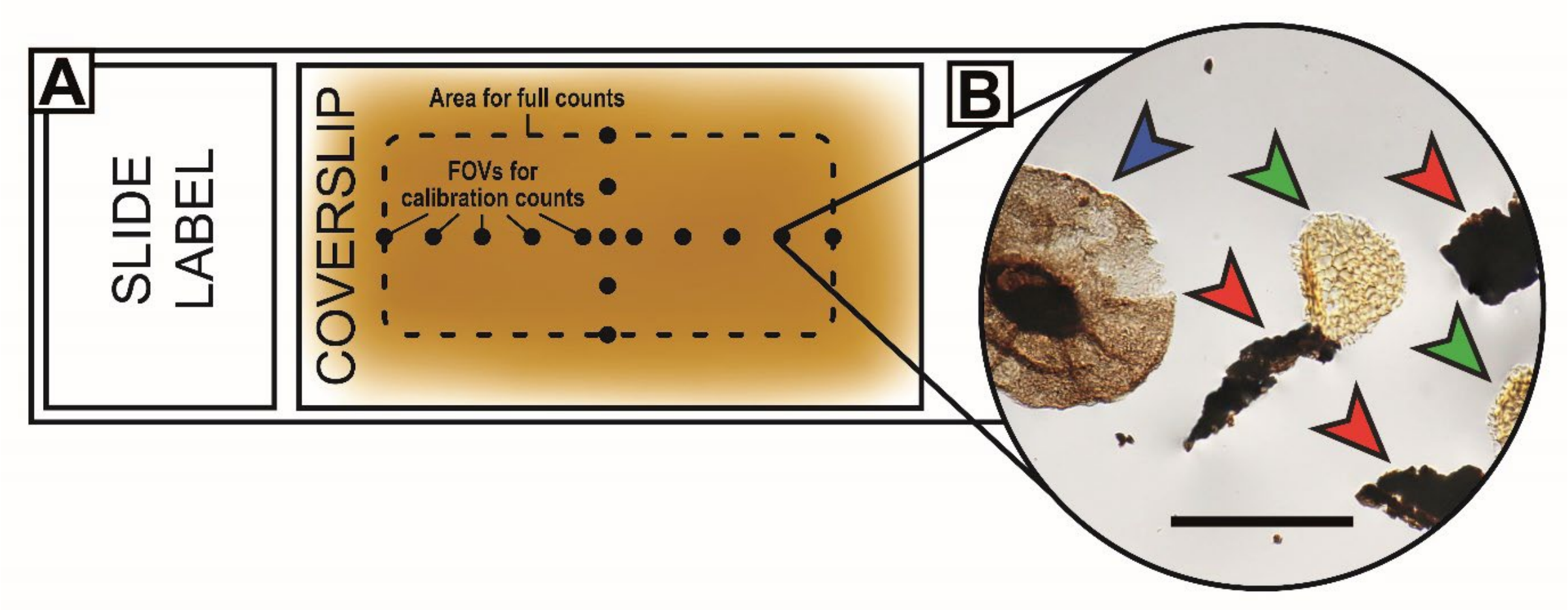

**Fig 2. Field-of-view subsampling (FOVS) method demonstrated with a schematic microscope slide of organic microfossils. A**, Organic microfossil slide, with specimens dispersed across a coverslip. For valid subsampling: 1, the spatial range of the full-count fields of view (FOVs) should match those of the calibration counts; 2, calibration counts should include both x- and y-axes; and 3, prevent edge effects by avoiding counts near the coverslip margins. **B**, A typical field of view of an organic microfossil (palynological) slide with potential targets and markers; red and blue arrows: terrestrial microfossils, fossil wood fragments (red) and a fossil plant spore, *Playfordiaspora crenulata* (blue); green arrows: exotic markers (modern *Lycopodium clavatum* spores); distinctive optical features of the markers are necessary for accurate and rapid data collection using the FOVS method; scale = 50 µm (sample S090320, Bonneys Plain-1, 301.56 m).

## Field-of-view subsampling (FOVS) method: precision estimates

While both absolute abundance methods (linear and FOVS) estimate the population values from subsamples, the accuracy of common specimen abundances using the FOVS method depends on the representativeness of the fields-of-views during the calibration counts. This entails a different source of error compared to the linear method. As such, the calculation of total error in Eqn 2 needs to be modified as follows:

$$\sigma_{Fx} = 100\sqrt{\left(\frac{s_{1P}}{\sqrt{N_1}}\right)^2 + \left(\frac{s_{3P}}{\sqrt{N_{3C}}}\right)^2 + \left(\frac{\sqrt{n}}{n}\right)^2} \quad (5)$$

where $\sigma_{Fx}$ is the total standard error of the concentration mean for the FOVS method (in %). In this function, $s_{1P}$ and $N_1$ are as in Eqn. 2, while $s_{3P}$ is the proportional sample standard deviation for the target specimens from the calibration counts and $N_{3C}$ is the number of fields of view during the calibration counts. For further calculations and descriptions of terms, see S2 Table.

Note that we have scaled $s_{3P}$ by the so-called $c_4$ correction, which is needed to ensure an unbiased estimation of the standard error of a sample mean with an underlying normal distribution (without this correction, the standard error is expected to understate the true value). This appears to be a classical result, with the earliest unambiguous reference we can find being [62] (eqn 16 therein). This $c_4$ correction is an artefact of the square root taken to obtain a standard deviation, and so if one uses the variance then it is not needed. However, standard deviation is more commonly cited in the literature, and so we utilise the more widely used standard deviation formulation in Eqn 5, but include the conservative correction in the $s_{3P}$ term (see the S1 Text).

Owing to their large sample sizes, standard sampling theory [58] tells us that the abundances of marker specimens per dose and the mean number of common specimens in the calibration-count fields of view ($N_{3C}$) both approximate Gaussian (or 'normal') distributions. Hence, their error functions—represented by the relationships between terms $s_{1P}$ and $N_1$, and $s_{3P}$ and $N_{3C}$, respectively—take similar forms in Eqn 5. This differs from the uncertainty associated with the target specimens ($x$) in Stockmarr's [25] total error estimate for the linear method (Eqn 2), where a single measurement for $x$ is taken and the statistics are assumed to have a Poisson distribution. However, given the typically low absolute abundances of the rare specimen ($n$) counts per field of view (often between 0.2 and 5), the uncertainty estimate associated with $n$ still likely follows a Poisson distribution. This type of distribution is particularly appropriate for discrete count data that are very rare (e.g., have a relatively low probability of occurrence within a given field of view; [63]). To demonstrate this, we present simulated data sets of rare specimen counts (mean number of rare specimens per field of view: 0.9, iterations: 100,000), and compare the probability distributions of two counts (Fig 3): 1, a single field of view ($N_{3C} = 1$); and 2, the sum of 15 fields of view ($N_{3C} = 15$). The red curves are the Poisson approximations, showing a very good visual fit, and the sum of squared residuals of $1x10^{-7}$ for the fit to the single field-of-view histogram and $1x10^{-5}$ for the total of 15 calibration-count fields of view. This allows us to conclude that the Poisson error estimate $\frac{\sqrt{n}}{n}$ for the rare grain type is a suitable approximation.

As in the case of the linear method, multiple target specimen types may be assigned for the FOVS method (e.g., $x_1, x_2, \dots, x_k$). For the FOVS method, however, the means per field of view ($\overline{Y}_{3,1}, \overline{Y}_{3,2}, \dots, \overline{Y}_{3,k}$) and standard deviations ($s_{3,1}, s_{3,2}, \dots, s_{3,k}$) of each target will need to be collected during the calibration counts. Combined with the markers ($n$) from the full counts, and following the sequence of steps outlined above (see 'field-of-view subsampling (FOVS) method: operation’), their individual concentration estimates could then be separately derived using Eqn 4. The number of appropriate targets is dependent on the relative abundances of these targets in the assemblage (i.e., the evenness of the assemblage). Should these additional targets be particularly rare, then their calibration counts will likely result in large standard deviations; hence, their predicted total error estimates will be correspondingly large.

It is predicted that for routine data collection effort (e.g., total counts of several hundred) the FOVS method will provide superior concentration precision (lower total error values) to the linear method for assemblages with suboptimal target-to-marker ratios (e.g., $\hat{u} > 3$). This is because in cases with disproportionately abundant common specimens, there will be diminishing returns in their additional counts [22], at which point, the key limiting factor in concentration precision will be the number of rare specimens. The power of the FOVS method lies in its ability to detect greater absolute abundances of rare specimens with minimal additional sampling effort, thus mitigating this issue. However, if the target relative abundance is exceedingly high—resulting in a very large target-to-marker ratio (e.g., $\hat{u} > 100$)—then it is predicted that not even the FOVS method will salvage satisfactorily precise concentrations without impractical sampling efforts, as the precision still relies on the absolute number of exotic markers counted. We test these predictions below by comparing the efficiencies of the FOVS and linear methods.

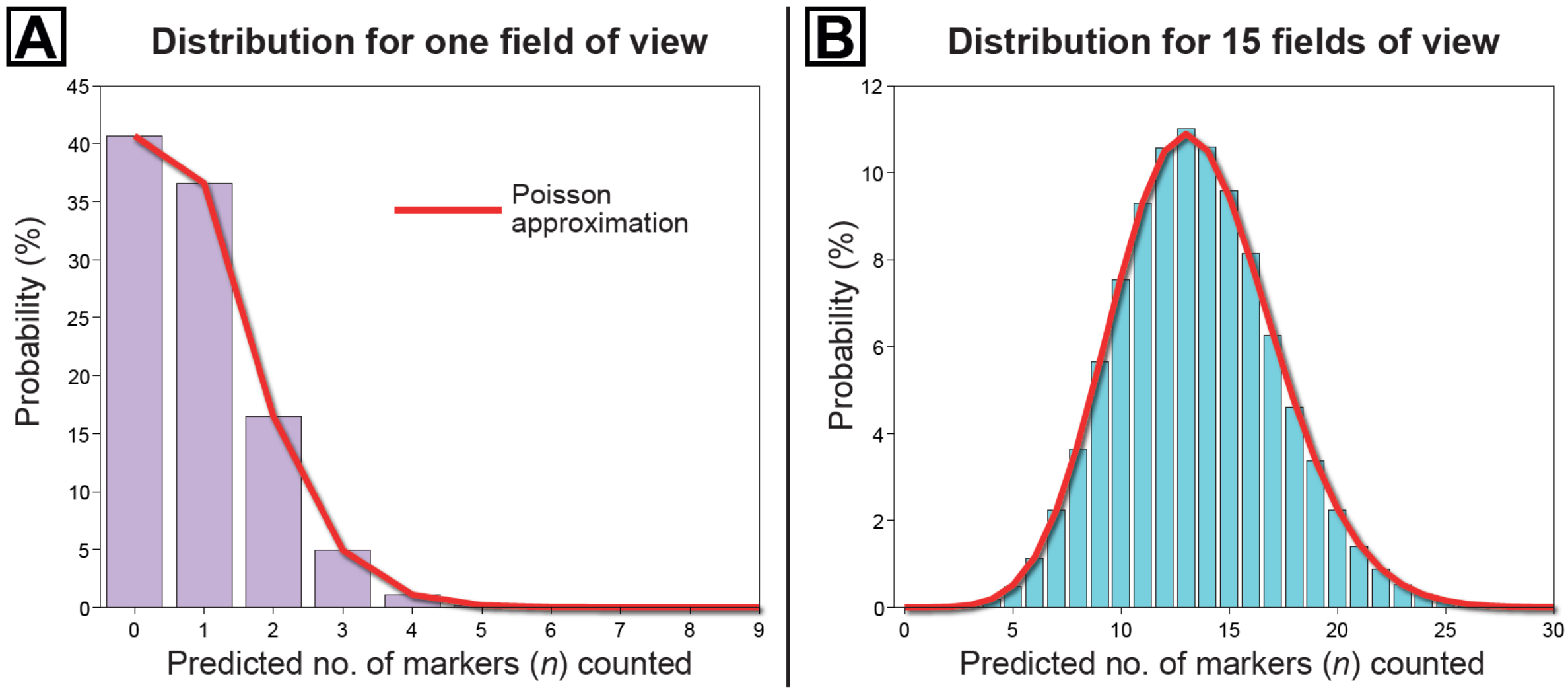

**Fig 3.** Histograms of counted marker specimens for an assemblage with rare markers; in this case, target-to-marker ratio [$\bar{\bar{u}}$] = 30; expected number of rare grains per field of view ($\overline{Y}_{3n}$) = 0.9; the number of simulations was 100,000. This is compared to the Poisson probability mass function with rate given by the actual mean number of markers per field of view calculated from the simulations. **A**, Histogram for a single random field of view. **B**, Histogram for the sum of 15 random fields of view.

## Estimating data collection effort

Given that the FOVS and linear method counts are conducted in subtly different ways, the amount of collection effort is not directly comparable. To standardise these comparisons, time was utilized as a proxy for data collection effort for both methods. The accurate identification and counting of target and marker specimens takes data collection effort (=time). Herein, we assumed an equivalent time required for the identification and counting of targets and markers. Moreover, the effort required for the calibration counts of the FOVS method is analogous to the counts in the linear method (for a given assemblage), since both require: 1, the same microscopy conditions (lighting, magnification, etc.); and, 2, the identification and counting of specimens on a given field-of-view before transitioning. In terms of effort, the primary difference is the addition of the subsequent 'full counts' for the FOVS method. These full counts typically entail a greater number of field-of-view transitions (i.e., the movements of the microscope stage to new fields-of-view), but fewer individual specimen identifications.

Sample collection effort is also a function of specimen density, which can vary greatly between assemblages. For example, the number of specimens on a given slide may be homogeneous but sparse, which would necessitate a larger number of field-of-view transitions, hence, a longer data collection time. Lower magnification fields of view would minimise the number of transitions, but decrease identification accuracy. The similarity between the FOVS method calibration counts and the linear method means that these two processes will be affected by specimen density to a similar degree. But the typically higher number of field-of-view transitions in the FOVS method means that this can represent a major component of the method's data collection effort, especially for low density assemblages.

For these reasons, a fair comparison requires that the numbers of examined fields of view be factored into the effort estimates of both methods. While field-of-view transitions may be extremely rapid (e.g., a few seconds on a given microscope slide), it may be generally slower than specimen counts within a single field of view. Microscopy trials counting total terrestrial organic microfossils yielded a time value for each field-of-view transition as approximately twice that of counting a single specimen on one field-of-view. However, we recognise that field-of-view transition effort factor (here denoted $\omega$, which is equal to the quotient of mean field-of-view transition time and mean specimen count time) and the identification of targets or markers may vary greatly between researchers and the types of targets being sought. For example, differentiating between visually similar targets to ensure accurate identification may inflate the specimen count time.

With all this in mind, the data collection effort for the linear method ($e_L$) was calculated as follows:

$$e_L = \left(\omega \times \frac{x}{\overline{Y}_{3x}}\right) + x + n, \quad (6)$$

where $\omega$ is the field-of-view transition effort factor, $x$ is the counted number of target specimens, $n$ is the counted number of marker specimens and $\overline{Y}_{3x}$ is the mean number of targets per field of view. For the microfossil case study herein, $\overline{Y}_{3x}$ was collected during the FOVS method calibration counts from the same slides of each sample. However, if linear method effort were to be estimated without this, the number of fields of view would need to be counted and substituted for the term $\frac{x}{\overline{Y}_{3x}}$ in Eqn 6, analogous to $N_{3C}$ for the FOVS method (see below).

In contrast, data collection effort for the FOVS method ($e_F$) was calculated by the formula:

$$e_F = (\omega \times N_{3C}) + x + (\omega \times N_{3F}) + n. \quad (7)$$

In this formulation, the sum of the common (typically, target) specimen counts ($x$) and the calibration-count fields of view ($N_{3C}$) collectively represent the effort during the calibration counts. The effort of the full counts is represented by the sum of the last two terms: the number of full-count fields of view ($N_{3F}$) and the number of rare (typically marker) specimens ($n$).

## Case studies

The precisions and efficiencies of the two count methods were analysed with two case studies, each of which followed an independent paradigm: 1, computer simulation; and 2,

‘real-world’ microfossil assemblage data. Three outcome variables were examined and compared between the two counting methods: 1, absolute abundance (concentration; Eqns 1 or 4); 2, error (= the degree of uncertainty to which the true concentrations in an assemblage can be inferred from a count; Eqns 2 or 5); and 3, data collection effort (Eqns 6 or 7).

## Case study: Computer simulation

To compare the count methods and optimise the new variant (the ‘field-of-view subsampling method’, or FOVS method) for maximum precision, we used a Monte Carlo simulation of ‘virtual study areas’ containing targets and markers. Monte Carlo simulations are a method of generating and testing hypotheses for complex systems using a large number of random instances of the scenario being studied. They have been successfully used in many scientific and mathematical contexts; a small set of examples include agriculture (e.g., [64]), cell biology (e.g., [65]), condensed matter physics (e.g., [66]), finance (e.g., [67]) and number theory (e.g., [68]). The simulation parameters were designed to mimic the (approximately) random distributions of specimens on a virtual study area; e.g., microfossils on microscope slides. Microscope slides are a small, well-controlled sample of approximately random particles, which are time- and labour-intensive to analyse in the laboratory, and so they are ideal candidates for a Monte Carlo approach.

The purpose of this simulation was to produce a series of idealised data sets with which to refine and optimise the FOVS method, and then compare the precisions of the linear and FOVS methods for an equivalent amount of effort. For each iteration of the simulation, a fixed number of each specimen type was distributed uniformly at random on a square virtual study area (Fig 4A). Between each sequence of iterations, however, the numbers of each specimen type were varied to test for the effects of target-to-marker ratio ($\hat{u}$).

Another benefit of these Monte Carlo simulations is that for each virtual study area, we can specify exact specimen population sizes. This allows us to calculate quantities not possible in most real-world study areas. For example, both the simulation and empirical approaches (see "case study: Permian–Triassic organic microfossils of southeastern Australia") provide the means for comparing precision of the methods, reflected by their standard errors. However, the simulations can also measure the accuracy of each method's error estimates, indicated by variations of the error estimates from the known true values (see S5–S15 Tables, S17 Fig).

Since we are comparing the linear method to the FOVS method on the same data sets, the error contribution from the addition of marker doses ($T = {s_{1P}}^2/N_1$; e.g., *Lycopodium* spore tablets) was the same for both methods. So, for the simulations to provide the clearest comparisons of the two methods, we have specified the number of markers exactly. In other words, there was no error contribution from the exotic markers. This gives us a precise value of $N_1 \times \overline{Y}_1$ in Eqns 1 and 4, and consequently results in $s_{1P} = 0$ (see Eqns 2 and 5). Additionally, we set the total sample dimension size ($\overline{\overline{V}}$) as 1 arbitrary size units; this parameter is analogous to the total mass or volume of a sediment from which microfossils might be counted.

Each virtual study area consisted of randomly distributed targets and markers of predetermined population sizes. We then applied the following procedures to each virtual study area.

*Linear method.* Starting at a fixed point on the virtual study area, a contiguous region of pre-determined height and variable length was marked out (Fig 4). The length was modified for each virtual study area to encompass a specified number of target specimens ($x$). This provided our random "window" within which the number of markers ($n$) were also counted. These data were then used to produce the following simulated parameters of the linear method: target specimen concentration ($c_L$; Eqn 1), standard error ($\sigma_L$; Eqn 2) and sampling effort ($e_L$; Eqn 6).

*FOVS method.* A target level of effort ($e_F$) was chosen to match the expected level of effort for the linear method for the same virtual study area. This target level of work was used to estimate the optimal numbers of calibration fields of view ($N_{3C}^*$) and full-count fields of view ($N_{3F}^*$) for the FOVS method (see "optimising the FOVS method"). Then the total number of common specimens were counted from this optimal number of calibration-count fields of view, from which we calculated the sample mean ($\overline{Y}_3$) and proportional sample standard deviation ($s_{3P}$) of the common specimens (note: in all simulated cases, the number of targets was equal to or greater than markers [$x \geq n$]; so, the targets [$x$] were the foci of the calibration counts, and $\overline{Y}_3 = \overline{Y}_{3x}$.) Next, the number of rare specimens (= markers [$n$] in all simulated cases) were counted from the corresponding optimal number of full-count fields of view ($N_{3F}^*$). With these data, we then calculated the following parameters for the FOVS method: target specimen concentration ($c_{Fx}$; Eqn 4), standard error ($\sigma_{Fx}$; Eqn 5) and the sampling effort ($e_F$; Eqn 7).

Note that we experimented with the parameters for the linear method, so that the level of work required for the FOVS method kept the total number of fields of view to a suitably small number, specifically $1 \leq N_{3C} + N_{3F} \leq 361$. (Our simulated virtual study areas can have a maximum of 1089 non-overlapping fields of view.) By keeping the maximum sampled area much smaller than the total study area, we can avoid significant “finite population effects”, whereby too many samples from a finite area will produce distorted statistical estimates. In the extreme case, if the total sample count contains the entire population, then the sample standard deviation will be zero. To ensure we make precise comparisons between the estimates from the two methods (linear and FOVS) with the computer simulations, we included a “finite population correction” (FPC; Scheaffer *et al.*, 2012, p. 83; see S1 Text).

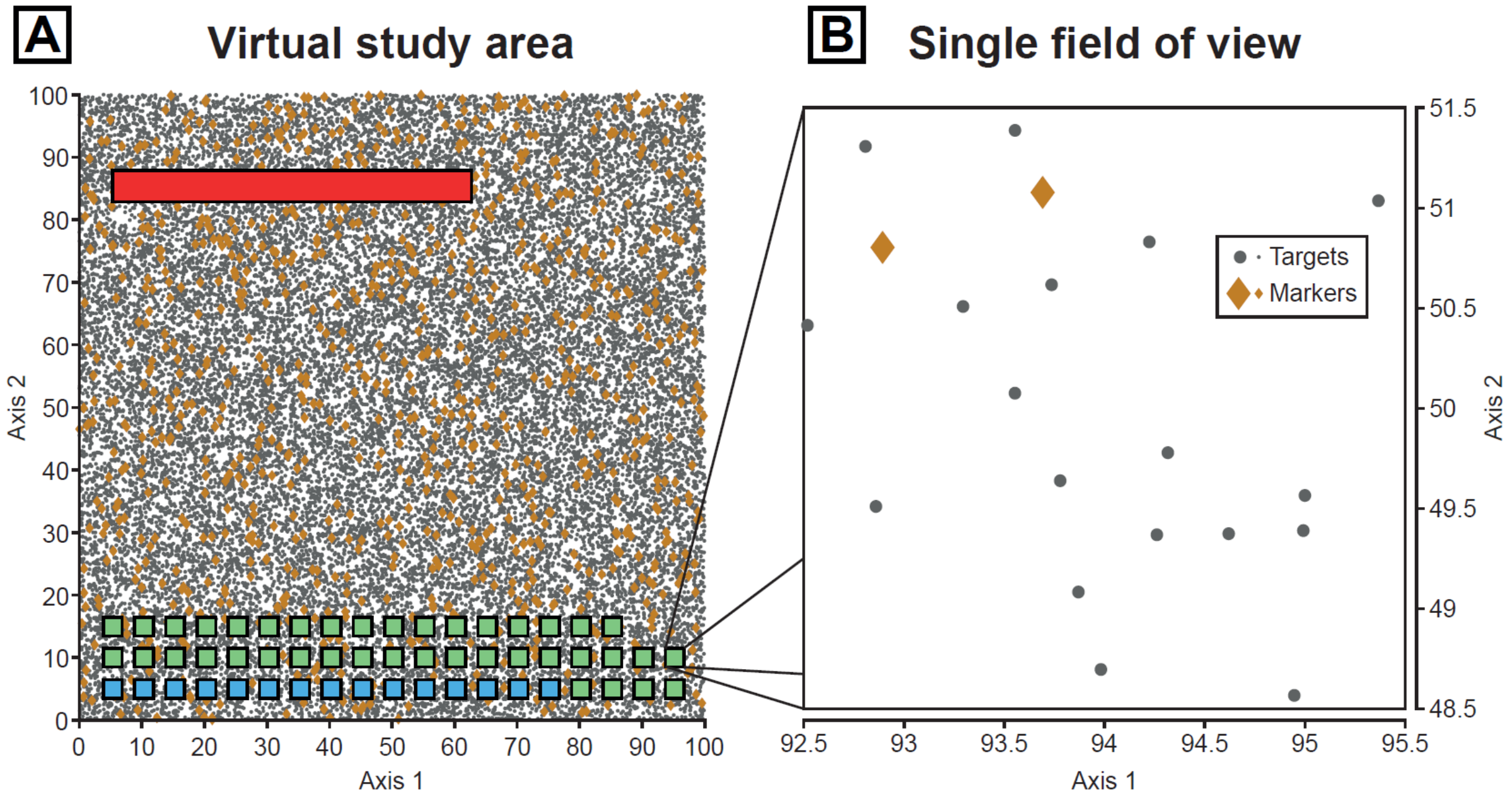

**Fig 4. Visualisations of a virtual study area to test the precision of the linear vs FOVS methods for a given amount of collection effort.** It includes the random distribution of 30,000 target specimens (grey dots) and 1,000 markers (brown diamonds); hence, $\bar{\bar{u}}$ = 30. **A**, Full virtual study area, blue squares are the individual fields of view used for the FOVS method calibration counts, red rectangle is the "window" used for the linear method

data, green squares are examples of the full-count fields of view. **B**, A representative field of view for the full or calibration counts used in the FOVS method.

*Simulated data sets.* For each set of parameters, we ran the simulation 1,000,000 times, producing data sets of the concentration, total error and effort for each iteration of both methods (see S5–S15 Tables). We then computed the mean values of: 1, the estimated concentrations ($c$; Eqn 1); 2, the total errors ($\sigma_L$ and $\sigma_F$; Eqns 2 and 5), recalling that we have no error contribution from the marker doses; 3, the standard deviation from the known exact concentrations ($\sigma_{exact,M}$; S18 Eqn); and 4, the required data collection effort ($e_L$ and $e_F$; Eqns 6 and 7). For maximum precision of the simulations, additional statistical corrections and effort standardisation across the different methods were included (see S1 Text).

The simulations were generated in Matlab v.R2020a. We have provided the code to generate these simulations (S16 Text), which can be downloaded here: https://github.com/Palaeomays/FOVS_vs_linear_methods.git. For each instance of the simulation, the user chooses:

- the number of targets and markers on the slide;
- the error contribution from the marker doses;
- the number of effort units;
- the field of view transition factor ($\omega$); and
- the number of iterations of the simulation.

## Case study: Permian–Triassic organic microfossils of southeastern Australia

*Processing, imaging and sample details.* All empirical data analysed herein derive from one drill core within the Tasmania Basin, southeastern Australia: Bonneys Plain-1 (41° 46' 27.69"S, 147° 36' 13.35"E; Fig 5). The target strata were part of the upper Parmeener Supergroup and, although these strata lack precise age control, they correlate to the upper Permian (Lopingian) to Lower Triassic Series [69]. For organic microfossil processing and comparisons of linear and FOVS count methods, 18 samples were chosen at random stratigraphic heights throughout the drill core (see S4 Table for details). The assemblages consisted of a range of shallow marine to coastal plain palaeoenvironments ([69]; C.R. Fielding, pers. comm.). Inorganic mineral content was removed by digestion with hydrochloric acid followed by hydrofluoric acid. Prior to acidification (following [42]), a number of tablets of *Lycopodium clavatum* spores were added; these spores served as the exotic markers in this case study. Tablets were produced by the Department of Geology, University of Lund, Sweden. Details of the tablets (including means and uncertainty estimates of *Lycopodium* spores per tablet and batch numbers), the specific quantities of *Lycopodium* tablets ($N_1$) added to each sample and spore quantity estimates of each tablets ($\overline{Y}_1$, $s_1$, $s_m$) are provided in S4 Table. No sieving, heavy liquid separation or oxidation was performed for these 'kerogen' residues. The resultant residues were mounted on glass slides, and glass coverslips were sealed with epoxy. A summary of these various aspects of palynological processing techniques was provided by Riding [70]. These residues of 'sedimentary organic matter' (*sensu* [71]) are the target of palynofacies analysis (*sensu* [72]), and are reflective of the undissolved particulate organic carbon (*sensu* [73, 74]), of a sediment sample's total organic carbon.

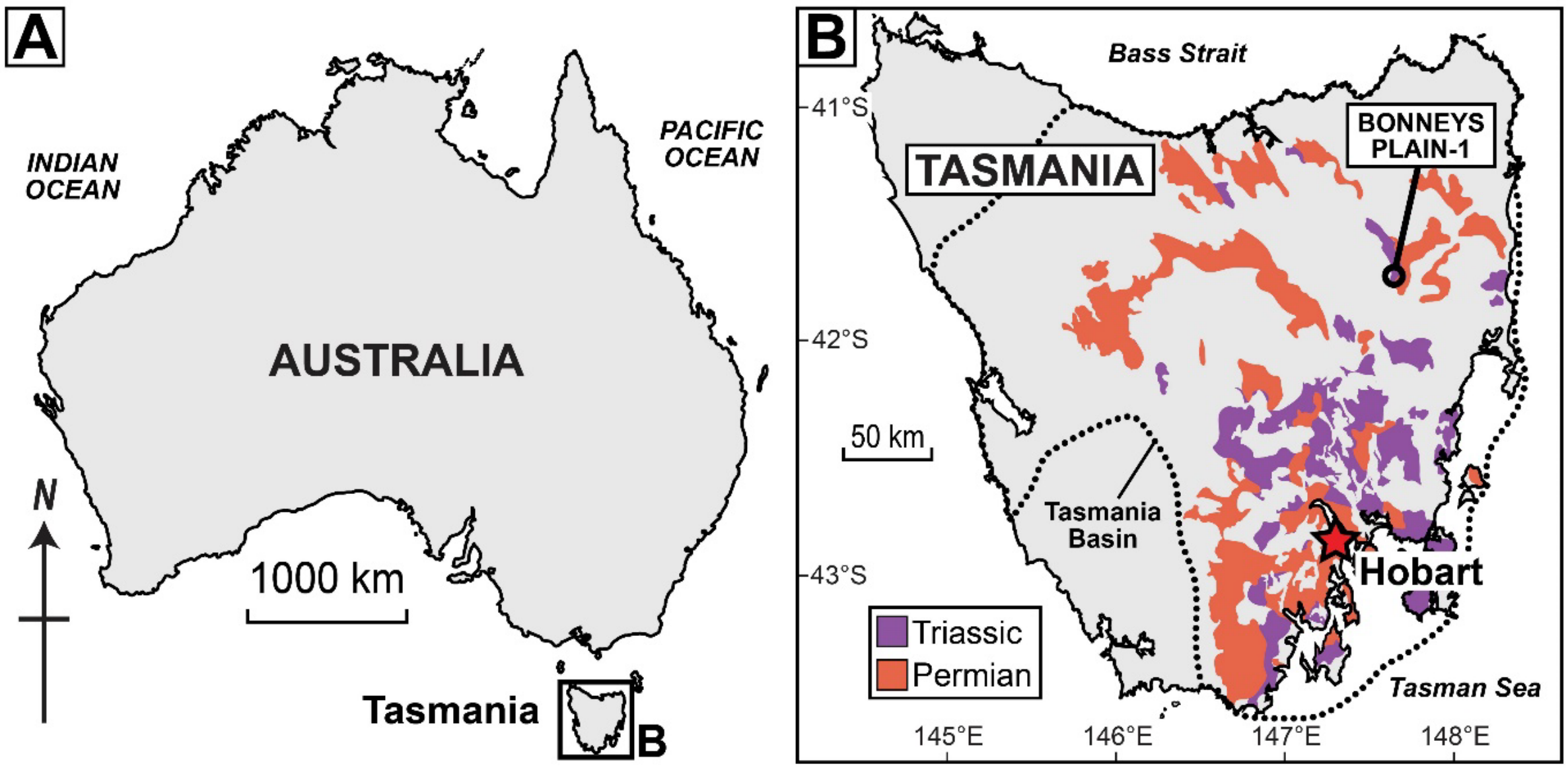

**Fig 5. Geographic and geologic contexts for the Permian–Triassic organic microfossil assemblages. A**, Map of Australia. **B**, Geological map of the Tasmania Basin [75] with location of target well core succession (Bonneys Plain-1) and approximate distributions of Permian-Triassic sedimentary strata.

By targeting mudrock facies of continental environments (e.g., lacustrine, fluvial, coastal), these microfossil assemblages were likely the result of low-energy, suspension deposition proximal to their sources (e.g., [76–80]). This is reflected by the high proportions of plant-derived phytoclasts in most assemblages. Moreover, estimating microfossil concentrations from sedimentary organic matter assemblages avoids sieving, oxidation and heavy liquid separation, each of which has potential for unintentional biasing effects ([40]; see reviews [70, 81]).

These samples were processed by Global Geolab Limited in Medicine Hat, Canada. Transmitted light microscopy and photomicrography of organic microfossils was conducted with a Zeiss Axioskop 2 transmitted light microscope equipped with a Canon EOS 700D camera. Fluorescence photomicrographs were performed with a Leica K3C camera on a Leica

DM2500 LED microscope. Residues and slides have been given the prefix 'S' and are housed at the Department of Palaeobiology, Naturhistoriska riksmuseet (NRM), Stockholm, Sweden. Additional specific methodological details for the different concentration estimates discussed are provided below where relevant.

*Count methods.* In these assemblages, the principal goal was to estimate the concentration of terrestrial organic microfossils ($c_t$). To this end, terrestrial organic microfossil grains were the target specimen type ($x$) and the subjects of the calibration counts; spores of *Lycopodium clavatum* were the exotic marker grains ($n$). The terrestrial organic microfossil populations consisted of (in order of approximate decreasing abundance): wood (='phytoclasts', including charcoalified wood), leaves, plant spores, pollen and fungal remains. Fossil resins and animal-derived clasts were negligible. For both count methods, only grains ≥5 µm in diameter were counted; specimens smaller than this could not be consistently identified.

The assemblages in this case study had the following criteria that made them particularly suitable to the FOVS method of data collection: 1, the *Lycopodium* spores were optically distinct from the other microfossils (e.g., colour, texture, transparency, fluorescence response), even from fossil spores or pollen (Figs 2B, 6); and 2, generally, one of these grain populations was relatively rare (mean target-to-marker ratio for Bonneys Plain-1 was high: $\hat{u} > 10$). All fields of view were examined with a 63× magnification objective. The field-of-view transition effort factor ($\omega$) was measured as approximately twice that of counting a single specimen on one field-of-view (i.e., $\omega = 2$). All assemblages included a sampling effort (see below) of >500. To minimise observer expectation biases, all counts followed the "blind protocol" outlined by Mays & McLoughlin ([21], p. 297); specifically: 1, all slide labels were masked; 2, slide order was randomized; and 3, sample counts were then conducted.

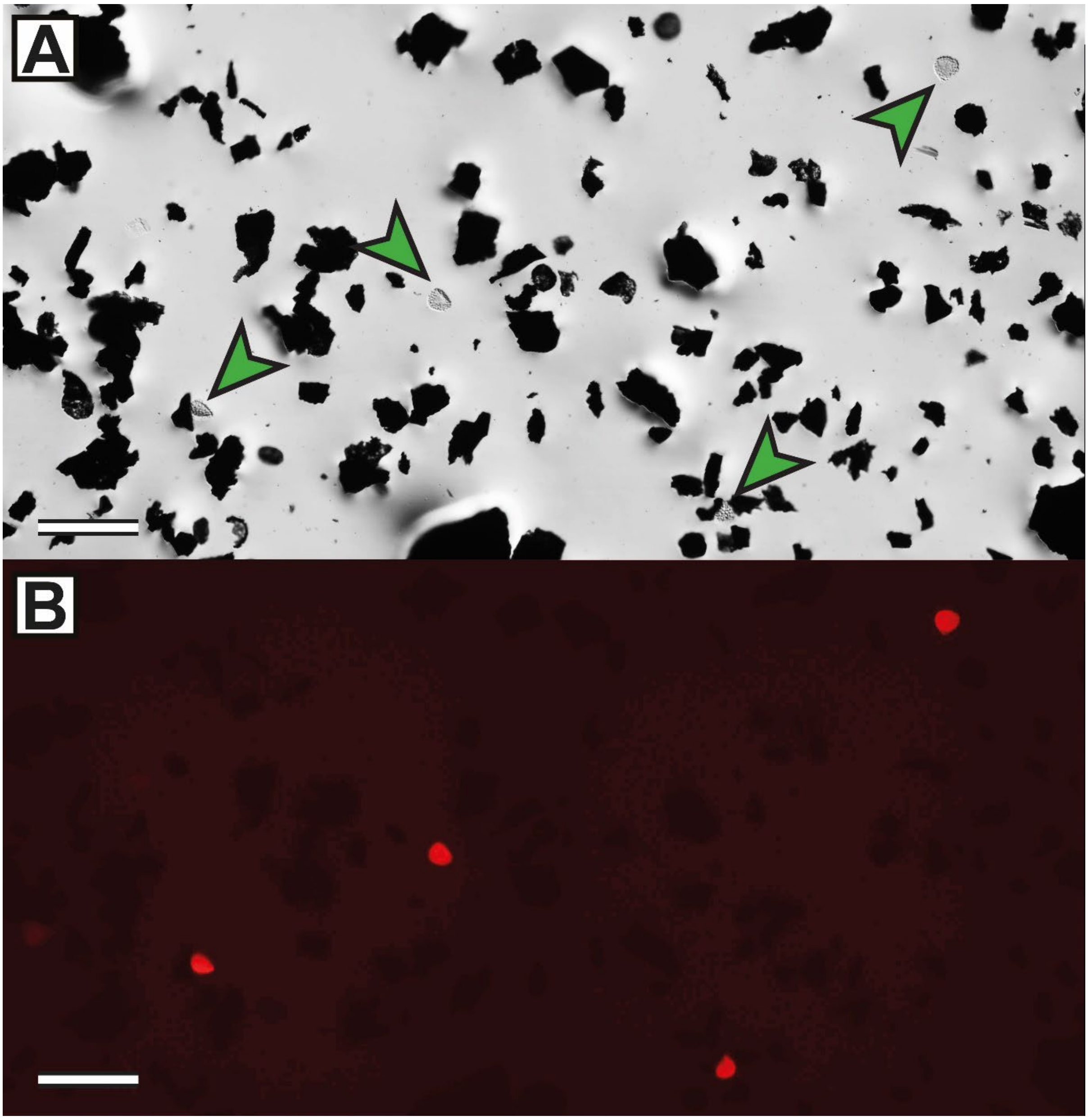

**Fig 6. A typical field-of-view photomicrograph of a palynological assemblage spiked with *Lycopodium* spore markers.** Specimen S090248, well core MPT-3, Tasmania Basin; scales = 100 µm. **A**, Greyscale optical light microscopy (with differential interference contrast). **B**, Fluorescence microscopy (excitation wavelength: 555 nm); note the strong autofluorescence of the *Lycopodium* markers (green arrows).

## Simulating precision as a function of data collection effort

The simulations enable us to define the assemblage characteristics precisely, against which we can test the efficiency of each method by how close these methods reach the predefined values. Hence, in the simulated case study, we can denote these pre-defined 'true' values of targets ($\bar{\bar{x}}$) and markers ($\bar{\bar{n}}$), with the population ratio of target to marker specimens ($\bar{\bar{u}}$) as

$$\bar{\bar{u}} = \frac{\bar{\bar{x}}}{\bar{\bar{n}}} \geq 1, \tag{8}$$

which reflects the typical scenario whereby the target specimens are more common than the exotic markers (following Eqn 3). While $\bar{\bar{u}}$ is known in the simulations, the sample target-to-marker ratio ($\hat{u}$) is an approximation of this value; this sample variable is given by: 1, $\hat{u} = x/n$ for the linear method; or 2, $\hat{u} = \overline{Y}_{3x}/\overline{Y}_{3n}$ for the FOVS method, where $\overline{Y}_{3x}$ is the mean number of targets per field of view, and $\overline{Y}_{3n}$ is the mean number of markers per field of view.

### Linear method

For the linear method, we note that the expected amount of effort is directly proportional to the number of specimens that we count in a window. To see this, we substitute $n = \hat{u}x$ into Eqn 6 to measure linear method effort, thus obtaining

$$e_L = \left(\frac{\omega}{\overline{Y}_{3x}} + 1 + \frac{1}{\hat{u}}\right)x = Ax, \tag{9}$$

We can then substitute $A$ (which reflects the degree of effort for each specimen in the linear method) into Eqn 2 to give us an expression for the precision as a function of the amount of effort:

$$\sigma_L(e_L) = 100\sqrt{T + (1+\hat{u})\frac{A}{e_L}}, \tag{10}$$

where $T$ is the error contribution from the marker doses ($T = {s_{1P}}^2/N_1$). As we increase the data collection effort ($e_L$), we naturally expect an increase in precision (i.e., a decrease in error, $\sigma_L$). Our derivations of Eqn 10 (see the 'analysis of precision as a function of effort and the ratio of common and rare grains' section of S1 Text) reveal: 1, there are steeply diminishing returns from increased effort with the linear method, as error decreases with the square of the effort (see Eqn S1); and 2, that the highest precision estimates of the concentration are achieved when $\hat{u} = 1$ (i.e., $x = n$; Eqn S2), thus confirming the result by Regal & Cushing [55] mentioned in the introduction.

## FOVS method

We would like to obtain an estimate of the improvement in FOVS method precision ($\sigma_F$) for increased effort ($e_F$), as we did for the linear method. However, we cannot do this directly, since the FOVS method involves two independent variables that determine the amount of effort: 1, the number of calibration fields of view ($N_{3C}$), which increases $x$; and 2, the number of full-count fields of view ($N_{3F}$), which increases $n$. Yet, if we can find the optimal ratio of calibration and full-count fields of view, then we will reduce our two-dimensional problem to a single dimension. Ultimately, this was done by employing a standard optimisation technique: calculate the derivative of the FOVS method error function (Eqn 5) and set it equal to zero (see 'FOVS method optimisation' below).

To make this derivative tractable, however, we will assume that the error term for the common specimens (typically, this will be the target specimens) $\left(\frac{s_{3P}}{\sqrt{N_{3C}}}\right)^2$ is well-approximated by the Poisson result $\left(\frac{x}{\sqrt{x}}\right)^2$. Given that this approximation is quite good for small numbers (as shown in Fig 3), we expect that this Poisson assumption will not introduce any error into our determination of the optimal field-of-view count ratio, since any small deviations will be erased when the numbers of calibration-count and full-count fields of view are rounded off to the nearest integer. The assumption may introduce a small degree of error into the formula for choosing between the FOVS and linear methods (see 'choosing the superior count method'); however, our predictions for this choice match the data generated in the simulations (see S1 Text). Thus, we have confidence that the assumption is sound for this purpose, especially given the large differences in $\hat{u}$ values expected in real assemblages.

So, with the assumption of a Poisson distribution for the target grain error term, we make the replacement $x = N_{3C}\overline{Y}_{3x}$, using the definition of $\overline{Y}_{3x}$ in Eqn 3. (Note: $x$ is the number of targets counted during the calibration counts.) Also, using the estimate for the average density of marker grains in each full-count field of view $\overline{Y}_{3n} = \overline{Y}_{3x}/\hat{u}$, we make the following replacement: $n = N_{3F}\overline{Y}_{3n} = \frac{N_{3F}\overline{Y}_{3x}}{\hat{u}}$. Substituting these into Eqn 5 gives us

$$\sigma_F(N_{3C}, N_{3F}) = 100\sqrt{T + \frac{1}{N_{3C}\overline{Y}_{3x}} + \frac{\hat{u}}{N_{3F}\overline{Y}_{3x}}}, \qquad (11)$$

where we have included the arguments of the function to explicitly denote that this is a two-dimensional function, and to make the following calculations clearer.

From the results of the derivations (see S1 Text for details of this procedure), the increase in precision decreases with the square of the effort (S3 and S4 Eqns). This is analogous to the linear method (S1 Eqn) and, once again, the error depends in a non-trivial way on the target specimen density ($\overline{Y}_{3x}$) and the target-to-marker ratio ($\hat{u}$).

*FOVS method optimisation.* The FOVS method error depends in large part on the numbers of both calibration-count fields of view ($N_{3C}$) and full-count fields of view ($N_{3F}$). To find the optimal ratio of these counts, first we solve Eqn 7 for $N_{3F}$

$$N_{3F} = \frac{e_F - (\omega + \overline{Y}_{3x})N_{3C}}{\omega + (\overline{Y}_{3x}/\hat{u})} \tag{12}$$

and then insert this equation for $N_{3F}$ into Eqn 11 to obtain

$$\sigma_F(N_{3C}) = 100\sqrt{T + \frac{1}{N_{3C}\overline{Y}_{3x}} + \frac{\omega(\hat{u} + \overline{Y}_{3x})}{\overline{Y}_{3x}(e_F - [\omega + \overline{Y}_{3x}]N_{3C})}}. \tag{13}$$

For a fixed amount of effort, we can now find the number of calibration fields of view that minimises this error ($N_{3C}^*$). This is a two-dimensional optimisation problem, and is calculated by taking the derivative of $\sigma_F$ with respect to $N_{3C}$ (Eqn 13) and setting it to zero:

$$\frac{\partial}{\partial N_{3C}}\sigma_F(N_{3C}) = 0 \Rightarrow N_{3C}^*(e_F) = \frac{e_F}{\omega + \overline{Y}_{3x} + \sqrt{(\omega + \overline{Y}_{3x})(\omega\hat{u} + \overline{Y}_{3x})}}. \tag{14}$$

Substituting $N_{3C}^*(e_F)$ into Eqn 12, we can then find the corresponding optimal number of full-count fields of view ($N_{3F}^*$)

$$N_{3F}^*(e_F) = \frac{e_F\hat{u}}{\omega\hat{u} + \overline{Y}_{3x} + \sqrt{(\omega + \overline{Y}_{3x})(\omega\hat{u} + \overline{Y}_{3x})}}. \tag{15}$$

By dividing Eqn 15 by Eqn 14, we can provide the ratio of the optimal numbers of full-count ($N_{3F}^*$) and calibration-count ($N_{3C}^*$) fields of view. The sampling effort ($e_F$) cancels out, leaving us with the optimal field-of-view count ratio ($\delta^*$) in the following simple formula:

$$\delta^* = \frac{N_{3F}^*(e_F)}{N_{3C}^*(e_F)} = \hat{u}\sqrt{\frac{\omega + \overline{Y}_{3x}}{\omega\hat{u} + \overline{Y}_{3x}}}. \tag{16}$$

This is perhaps one of the most useful quantities for the practitioner of the FOVS method. It tells us that the primary variable in determining the most efficient number of field-of-view counts for a given assemblage is the ratio of target to marker specimens ($\hat{u}$). When $\hat{u} = 1$, we have $\delta^* = 1$, and so we should choose an equal number of full-count and calibration-count fields of view. However, in the more typical scenario where there are greater numbers of targets than markers ($\hat{u} > 1$), then the proportion of full-count fields of view should increase as the square root of $\hat{u}$.

The optimal field-of-view ratio ($\delta^*$) depends to a lesser degree on the density of target specimens across the study area ($\overline{Y}_{3x}$) and an observer's counting habits (reflected by $\omega$). Crucially, all three of these variables can be estimated during the calibration counts. To maximise the utility of the FOVS method for routine concentration estimates, we recommend utilising the relatively simple metric in Eqn 16 during the calibration counts for determining the optimal ratio of full to calibration counts. The optimal field-of-view count ratio can be calculated easily by inserting the relevant variables directly into the user-friendly interface we have provided (link here: https://github.com/Palaeomays/FOVS_vs_linear_methods.git; additional information in S1 Text). Note that the optimal ratio of calibration to full counts will not tell a practitioner how many of each should be counted for a given precision; this is discussed below (see 'achieving a targeted precision').

Lastly, if we wish to validly compare the efficiency of the two methods, we can now utilize the optimal field-of-view counts to characterise the FOVS method's relationship between error vs effort. Substituting the optimal values $N^*_{3F}(e_F)$ and $N^*_{3C}(e_F)$ into Eqn 11 gives the FOVS method precision as a function of the effort

$$\sigma_F(e_F) = 100\sqrt{T + \frac{1}{e_F\overline{Y}_{3x}}\left([1+\hat{u}]\omega + \overline{Y}_{3x} + \sqrt{[\omega + \overline{Y}_{3x}][\omega\hat{u} + \overline{Y}_{3x}]}\right)}, \qquad (17)$$

which is analogous to the linear method error vs effort function in Eqn 10. In the supplementary material (see “analysis of precision as a function of effort and the ratio of common and rare grains”), we used S1 and S5 Eqns to show that the FOVS and linear methods both have the same "one on error-squared" decrease in error for increasing effort. So, we need to analyse the methods more closely to determine which is superior; these analyses are provided in the following section.

## Choosing the superior count method

The quality of a data collection method is measured by both its accuracy and its efficiency, the latter of which is the result of two competing variables: precision ($\sigma$) and sampling effort ($e$). Having optimised the ratio of calibration and full count fields of view for the FOVS method, we have now established standardised metrics of effort and error (=the inverse of precision) for both methods (Eqns 10 and 17). So, we can now directly compare the errors of the linear and FOVS methods for the same amount of effort ($e$). This will give the reader a formula for determining which method is the most efficient choice for their own studies.

By taking the ratio of the errors (for the FOVS method error [$\sigma_F$], the optimal field-of-view counts [$N_{3C}^*$ and $N_{3F}^*$] are used), we obtain

$$\frac{\sigma_L}{\sigma_F} = \sqrt{\frac{\omega(\hat{u}+1) + (2+Te+\hat{u})\overline{Y}_{3x} + \left(\overline{Y}_{3x}/\hat{u}\right)}{\omega(\hat{u}+1) + (2+Te)\overline{Y}_{3x} + 2\sqrt{\left(\overline{Y}_{3x}+\omega\right)\left(\overline{Y}_{3x}+\hat{u}\omega\right)}}}. \tag{18}$$

If the value of this ratio is larger than one, then the FOVS method is expected to provide a smaller error for a given quantity of work, while a value smaller than one would indicate that the linear method is superior. The point at which the ratio equals one is given by the solution to the following equation:

$$(\hat{u}^2-1)^2\overline{Y}_{3x}^{\;2}-4\omega\hat{u}^2(1+\hat{u})\overline{Y}_{3x}-4\omega^2\hat{u}^3=0. \tag{19}$$

While this equation defines the dividing line between the choice of the two methods, the reader need not solve it. Instead, we can provide a simple formula for determining the appropriate method to use. The only parameters required for this determination are the assemblage-specific target-to-marker ratio estimate ($\hat{u}$) and the researcher-specific time parameter $\omega$ (the latter of which encompasses the time taken for field-of-view transitions and individual specimen counts). Given these data, we can then conduct a 'method determination test'. The parameter below provides the minimum density of target specimens per field of view for which the FOVS method is the superior choice ($\overline{Y}_{3x}^{*}$):

$$\overline{Y}_{3x}^{*}=2\omega\frac{\hat{u}^2+\sqrt{\hat{u}^3(1+\hat{u}[\hat{u}-1])}}{(\hat{u}+1)(\hat{u}-1)^2}. \tag{20}$$

So, if the mean number of target (or common) specimens per field of view for a given assemblage is larger than this number (i.e., $\overline{Y}_{3x}>\overline{Y}_{3x}^{*}$), then the FOVS method should be used; if it is smaller (i.e., $\overline{Y}_{3x}<\overline{Y}_{3x}^{*}$), then the linear method should be the preferred choice. Put another way, when specimen densities are low, a larger number of fields of view are needed to collect an accurate and precise data set, the time-cost of which disproportionately penalises the FOVS method. (Note: the subscript "$x$" for $\overline{Y}_{3x}^{*}$ and $\overline{Y}_{3x}$ denotes that the target specimens are the foci of the calibration counts; however, if the markers are the calibration count subjects, use the alternative equations in S1 Text).

## Achieving a targeted precision

Once we have settled on our method, we would like to obtain an estimate of the amount of effort required to achieve a desired maximum level of error. This desired error is denoted $\bar{\sigma}$ (expressed in %).

For the linear method, we simply replace $\sigma_L$ with $\bar{\sigma}$ in Eqn 10 and then rearrange it to find

$$e_L(\bar{\sigma}) = \frac{\omega(1+\hat{u}) + \overline{Y}_{3x}(2+\hat{u}) + \overline{Y}_{3x}/\hat{u}}{\overline{Y}_{3x}([\bar{\sigma}/100]^2 - T)}. \tag{21}$$

If, however, we have settled on the FOVS method for a given assemblage, then we would likely want to know how many calibration and full-count fields of view would be required to achieve a targeted level of error. First, we start with Eqn 11 and then rewrite it using Eqn 16 to replace $N_{3F}^*$ as follows:

$$\sigma_F(N_{3C}^*, N_{3F}^*) = 100\sqrt{T + \frac{1}{N_{3C}^*\overline{Y}_{3x}} + \frac{1}{N_{3C}^*\overline{Y}_{3x}}\sqrt{\frac{(\omega\hat{u} + \overline{Y}_{3x})}{(\omega + \overline{Y}_{3x})}}}. \tag{22}$$

Using this equation, we solve for $N_{3C}^*$ to determine the optimal number of calibration-count fields of view for our desired error ($\bar{\sigma}$):

$$N_{3C}^*(\bar{\sigma}) = \frac{1}{(\bar{\sigma}/100)^2 - T}\left(\frac{\sqrt{[\overline{Y}_{3x} + \omega]} + \sqrt{[\overline{Y}_{3x} + \hat{u}\omega]}}{\overline{Y}_{3x}\sqrt{[\overline{Y}_{3x} + \omega]}}\right). \tag{23}$$

We can then use Eqn 16 again to obtain an expression for $N_{3F}^*$ for the optimal number of full-count fields of view for a given error ($\bar{\sigma}$):

$$N_{3F}^*(\bar{\sigma}) = \frac{\hat{u}}{(\bar{\sigma}/100)^2 - T}\left(\frac{\sqrt{[\overline{Y}_{3x} + \omega]} + \sqrt{[\overline{Y}_{3x} + \hat{u}\omega]}}{\overline{Y}_{3x}\sqrt{[\overline{Y}_{3x} + \hat{u}\omega]}}\right). \tag{24}$$

The above equations (Eqns 23 and 24) tell us the number of fields of view needed for a specific level of error. To arrive at a prediction for the amount of work required to achieve that error, we substitute Eqn 23 into Eqn 13 and rearrange to find:

$$e_F(\bar{\sigma}) = \frac{2\bar{Y}_{3x} + \omega(1+\hat{u}) + 2\sqrt{(\bar{Y}_{3x}+\omega)(\bar{Y}_{3x}+\hat{u}\omega)}}{\bar{Y}_{3x}([\bar{\sigma}/100]^2 - T)}, \tag{25}$$

which can be compared to the analogous expression for the linear method (Eqn 21). This comparison is done simply by calculating the difference between the two methods:

$$e_L(\bar{\sigma}) - e_F(\bar{\sigma}); \tag{26}$$

this gives us another way to compare the efficiencies of the two methods (in addition to the function expressed in Eqn 20). Negative values for Eqn 26 indicate that the linear method would be more efficient, while positive values indicate that the FOVS method should be used (Fig 7).

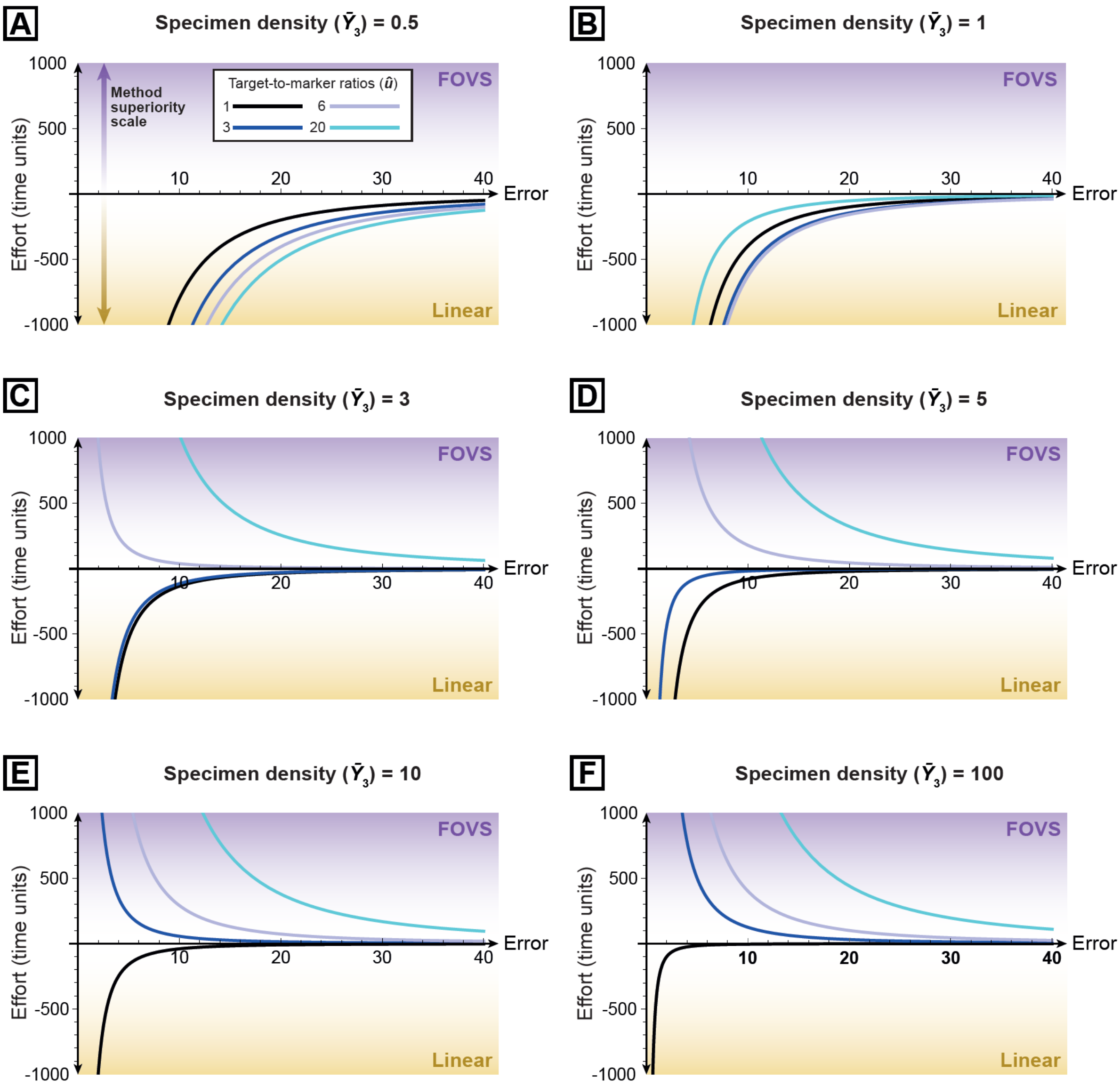

**Fig 7. Residual efficiencies of the linear vs FOVS data collection methods, and the influences of specimen density ($\overline{Y}_3$) and target-to-marker ratio ($\hat{u}$).** These plots illustrate the amount of net effort required for the linear method compared to the FOVS method (with optimal field-of-view counts; Eqn 26), where FOVS method data uses the optimal number of calibration- and full-count fields of view (Eqns 23 and 24). Hence, negative values indicate cases where the linear method has superior efficiency (with lower values indicating cases of even greater relative efficiencies for the linear method); the FOVS method is more efficient in all other cases. In all cases, error contribution from the

marker doses ($T$) is zero, and the ratio of field-of-view transition to specimen count time ($\omega$) = 2. **A**, $\overline{Y}_3 = 0.5$. **B**, $\overline{Y}_3 = 1$. **C**, $\overline{Y}_3 = 3$. **D**, $\overline{Y}_3 = 5$. **E**, $\overline{Y}_3 = 10$. **F**, $\overline{Y}_3 = 100$.

Of the variables included in Eqn 25, we can see that both target-to-marker ratio ($\hat{u}$) and target specimen density ($\overline{Y}_3$) have major roles in determining the most appropriate method. $\overline{Y}_3$ has a disproportionately large impact on the efficiency of the FOVS method; for assemblages with extremely low target densities (hence, very small $\overline{Y}_{3x}$), the linear method is always the best choice (Fig 7C), otherwise too much time is spent on field-of-view transitions. In the case of microfossils, however, assemblages with such low densities are exceptionally rare, and indicative of very poor fossil recovery. For assemblages with more reasonable (moderate to high) target specimen densities, then the linear method is only the best choice when there is roughly the same proportion of targets and markers (i.e., $\hat{u} \approx 1$; Fig 7A, B; S17 Fig), and even then, only marginally. In all other cases, the FOVS method is superior.

As noted above (see 'choosing the superior count method'), the most efficient method can be determined with very few input parameters (Eqn 20). Eqn 26 also provides this determination, while quantifying how much more efficient that method is over the other (for a given precision; Fig 7). Should the FOVS method be the superior choice, Eqns 23 and 24 provide the required number of fields of view for the calibration counts and full counts, respectively, for a desired precision level. For routine data collection, these equations will inform users about the feasibility of achieving satisfactory precision for each assemblage.

# Results

## Case study: Computer simulations

The simulated error vs effort relationships (summarised in Fig 8) demonstrate the following:

1) For both methods, a target-to-marker ratio ($\hat{u}$) of 1 produces the lowest error.
2) With near-equivalent numbers of markers and targets ($\hat{u} \approx 1$), both FOVs and linear methods require similar efforts to achieve errors of <8% (excluding error associated with the introduction of exotic markers; e.g., quantities of *Lycopodium* spores per tablet).
3) At $\hat{u}$ values close to 1, the linear method is slightly more efficient. However, even with a moderate difference between an assemblage's target and marker abundances—especially $\hat{u} \geq 3$—the amount of effort required for the linear method is consistently higher than the FOVS method for the same degree of precision (regardless of the FOVS method calibration count size).
4) In addition to the principal role that $\hat{u}$ plays in determining method choice, there is a non-trivial dependence on the mean density of targets in a field of view ($\overline{Y}_{3x}$). At moderate to high values of $\overline{Y}_{3x}$, the FOVS method is more efficient; at extremely low $\overline{Y}_{3x}$, the linear method is superior (all other variables being equal).
5) The simulated case study enabled exact concentrations and errors to be calculated, against which the estimated total errors of both linear and FOVS methods could be compared (S1 Text; S5–S15 Tables). From these comparisons, it was clearly demonstrated that the error estimates for the FOVS method are extremely accurate, regardless of $\hat{u}$, and consistently superior to the linear method; the linear method precision estimates, in contrast, are particularly unreliable at high $\hat{u}$ values (S1B Fig).

6) The minor discrepancies in total error from the three different FOVS scenarios are largely due to the small standard deviations derived from the calibration counts ($s_3$). This reflects the low spatial heterogeneity in the randomly distributed simulated assemblage. Under such scenarios, it is far more efficient to collect a smaller number of calibration counts. However, if the specimen density of the sample is particularly heterogeneous, then the error of these calibration counts will be proportionally high, which will then require a higher effort to achieve a desired precision. (A systematic investigation of spatial heterogeneity effects are beyond the scope of the present study.)

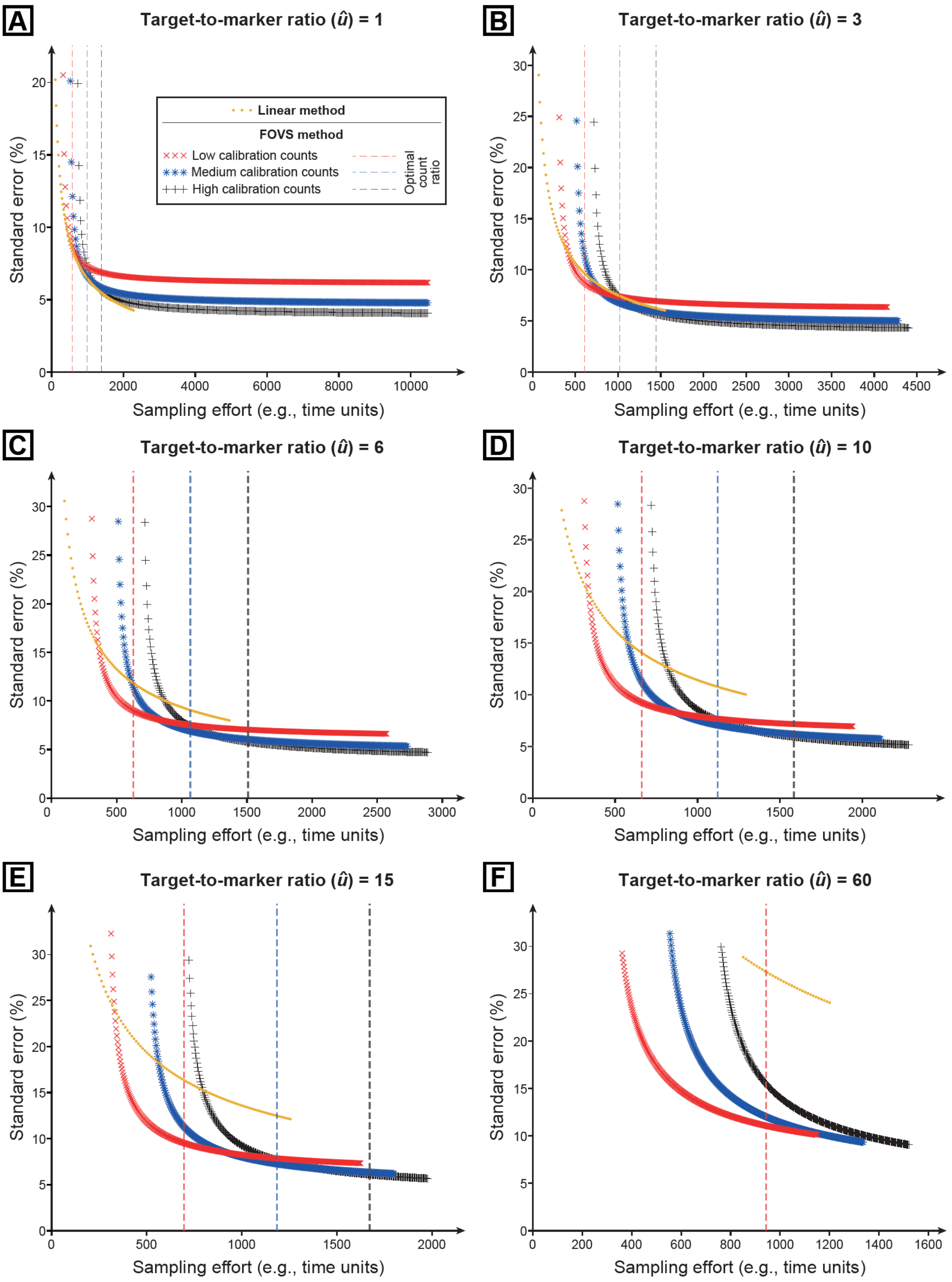

**Fig 8. Simulated data sets showing the relationships between precision and effort using the linear method (orange) vs the field-of-view subsampling (FOVS) method.** To test the effects of field-of-view count ratios, the three FOVS method plots represent three

different count conditions; each condition had a fixed number of calibration-count fields of view ($N_{3C}$), and effort was incrementally increased by sequentially adding full-count fields of view ($N_{3F}$); red: $N_{3C} = 10$ and $1 \leq N_{3F} \leq 351$; blue: $N_{3C} = 14$ and $1 \leq N_{3F} \leq 344$; black: $N_{3C} = 21$ and $1 \leq N_{3F} \leq 337$. The dashed lines indicate the optimum ratio ($\delta^*$ in Eqn 16) of full- to calibration-count fields of view for each of the three FOVS count conditions. (Note that for $\bar{\bar{u}} = 60$, only the red dashed line appears, as there were not enough full-count fields of view in this simulation to reach the optimal ratio for the other two [blue and black] count conditions). For the linear method, effort was gradually increased by lengthening the "window" until a predetermined number of targets was counted. Error contributions from the markers have been set to zero ($T = 0$); hence, absolute errors will be underestimates of those expected in a natural sample, but the relative errors between methods will be accurate. Each simulated scenario underwent 100,000 iterations. Six different scenarios are provided, which reflect different target-to-marker ratios ($\bar{\bar{u}}$). **A**, $\bar{\bar{u}} = 1$; **B**, $\bar{\bar{u}} = 3$; **C**, $\bar{\bar{u}} = 6$; **D**, $\bar{\bar{u}} = 10$; **E**, $\bar{\bar{u}} = 15$; and **F**, $\bar{\bar{u}} = 60$.

## Case study: Microfossil data

When the linear and FOVS methods were applied to the Permian–Triassic terrestrial organic microfossil assemblages of the Tasmania Basin, the concentration estimates ($c_t$) produced by the two methods were very similar, with the FOVS method tending towards slightly higher values (Table 2). A paired $t$-test was used to test whether the differences in mean $c_t$ between these two methods were statistically significantly. The paired test was essential since the two populations consisted of random samplings using different count methods, but taken from the same populations (organic residues of the same sedimentary rock sample);

the null hypothesis predicted no difference in $c_t$. The paired $t$-test revealed a non-significant difference between the mean $c_t$ values for these methods ($t$ = 1.893; $p$ = 0.08; $N$ = 18). An illustration of this concordance: in all assemblages, the concentration estimates from the FOVS method counts fell within the 95% confidence intervals of the linear method counts (S2 Table).

The differences in precision between the two methods, however, were pronounced, as reflected by their total error values ($\sigma_L$ and $\sigma_F$ ). A Wilcoxon signed-rank test was employed to test whether the differences in median total error between the two methods were statistically significant. This non-parametric test was chosen because it applies to paired samples, but does not assume normally distributed data [82, 83]. The null hypothesis predicts zero differences between the medians. The FOVS method resulted in a much lower total error; this difference was highly statistically significant ($W$ = 171, $p$ = 0.0002; $N$ = 18). In principle, this high significance value is typical of: 1, a modest variation in precision (i.e., a modest 'effect size') for a large sample set; or, 2, a major effect size from only a small sample set. Since there was only a small number of assemblages in the comparisons ($N$ = 18), this indicates a major 'effect size'; specifically, a major improvement in precision for the FOVS method. Crucially, this precision improvement was despite a similar mean sampling effort (Table 2). Given that the two methods were conducted on identical samples, the calculated difference in mean concentration estimates (c. 6%) was likely due in large part to the relatively large error estimates of the linear method. Lastly, we utilised Eqns 23 and 24 ($e_L[\bar{\sigma}]$ and $e_F[\bar{\sigma}]$, respectively) to predict the sampling efforts required by the two methods to achieve the same error rate. To achieve an error of 10%, the linear method would require approximately 7 times the sampling effort than the FOVS method (Table 2).

| Parameter | Linear method | Field-of-view subsampling (FOVS) method |
|---|---|---|
| Concentration of terrestrial organic microfossils ($c_t$, grains/g), mean | 328478* | 344341* |
| Total standard error ($\sigma$, %), mean | ± 29.1** | ± 11.65** |
| Sampling effort ($e$), mean | 990 | 1064 |
| Calculated target-to-marker ratio ($\hat{u}$), post-count mean | 89.8 | 94.3 |
| Predicted effort ($e[\bar{\sigma}]$, in time units) for a target of 10% error, post-count mean | 10069 | 1445 |

**Table 2.** Comparison table of terrestrial organic microfossil concentrations ($c_t$), and their associated uncertainties (total standard errors) and sampling efforts. Samples (*N* = 18) are from the Permian–Triassic assemblages of Bonneys Plain-1, Tasmania Basin, Australia (see S4 Table). * = no statistically significant difference ($p > 0.05$); ** = highly statistically significant difference ($p < 0.0005$).

# Discussion

## Absolute abundance data collection techniques

The two methods employed herein yielded similar absolute abundances (or concentrations) in both simulated and empirical data sets. However, the quality of a new method is also demonstrated by its ability to produce the same or higher precision, for the same or lower sampling effort. Moreover, the superior method should provide a more accurate approximation of its precision. Despite the benefits of the 'field-of-view subsampling method' (FOVS method), including greatly decreased error rate with consistent sampling effort, there are important limitations to this technique that need to be balanced against the strengths and limitations of the linear method. The limitations of each method are expanded below and summarised in Table 3.

## Linear method: limitations

The key limitation of the linear method lies in its high sensitivity to the ratio of target specimens to introduced markers; i.e., the 'target-to-marker ratio' ($\hat{u}$). Regardless of the count type, the highest precisions are achieved when the target-to-marker ratio is close to 1:1 [22, 55]. We have demonstrated that, in these optimal conditions (and with reasonable values for other relevant parameters), the linear method shows a marginally superior efficiency (Figs 7 and 8). However, the linear method has a low tolerance for ratio values far beyond this [40, 42, 84]. This has important practical limitations:

1. <u>Reprocessing of samples.</u> For the microfossil assemblages discussed herein, the exotic markers are introduced during sample preparation; hence, before observation at the microscope [70]. However, it is impossible to accurately predict *a priori* how many markers to add without examining the prepared microfossil assemblages for approximate concentrations of the target specimens. Attempts to circumvent this apparent paradox typically involve undertaking two (or more) preparation phases. For instance, the preparator might: 1, predict the optimal number of markers to introduce for a given sample (often informed by sedimentary indicators of microfossil concentrations, e.g., sediment grain-size or colour, inferred depositional conditions, total organic carbon analyses); 2, conduct a preliminary batch of processing; 3, examine a subsample of the assemblage to gauge the approximate target-to-marker ratio; then, if the ratio is suboptimal, 4, conduct a subsequent phase of processing with a modified number of introduced markers. This may entail several processing rounds. Repeated processing can be costly in time, money and sediment (or sedimentary rock) sample, the latter of which may be irreplaceable; hence, this may be impractical for many potential applications. In principle, processing costs might be mitigated by adding the markers towards the end of the preparation procedure, as this would reduce the number of steps during reprocessing (e.g., acidification, oxidation; [40, 41]). However, the late-stage introduction of exotic markers can lead to their overrepresentation, because the early stages of sample preparation demonstrably bias the assemblages [42]. Hence, marker grains should be introduced at the start of the preparation process, resulting in a consistent degree of bias for both markers and targets [42], but also leading to a longer and more costly preparation should the sample need to be processed multiple times.

2. High data collection effort. In cases with suboptimal target-to-marker ratios, the key limiting factor in concentration precision will be the low abundances of the rare specimens (typically, these will be the exotic markers). Thus, increasing the number of the rare specimens counted, even by a modest amount, can provide an enormous increase in precision. The 'brute force' way to bolster the count of these is simply to undertake a longer data collection phase. However, since precision is the result of multiple factors (Eqns 2 and 5), the relationship between target-to-marker ratios and effort is nonlinear. We have shown (S1 Eqn) that the increase in precision decreases with the square of the effort. Thus, assemblages with higher ratio require inordinate increases in specimen counts—therefore, data collection effort—to achieve the same low degree of uncertainty [22, 25] (Fig 8). For example, if seeking a total error of 10% with a target-to-marker ratio of 3:1 ($\hat{u} = 3$), a density ($\overline{Y}_{3x}$) of 10 targets per field of view, and assuming similar exotic marker errors ($T$) and transition-to-count ratios ($\omega$) as those employed in this study, one would need a total effort of c. 627 effort units (calculated from Eqn 21). However, with a target-to-marker ratio of 20:1 ($\hat{u} = 20$), but all other parameters being equal, an effort of c. 2682 is required to achieve the same degree of precision. Under such conditions, it is entirely reasonable to reprocess the samples. As stated by Maher (1981, p. 188): "it requires very little effort to add a few marker tablets to some sediment [samples]. It *does* require effort to increase the pollen counts."

3. Unsuitable for multiple concentration targets. A limitation related to the linear method's sensitivity to target-to-marker ratios is that an 'optimal' ratio assumes only one target population. However, there may be more than one target population (e.g., $x_1, x_2, \ldots, x_k$) for which absolute estimates are being sought, each of which will have its unique concentration value (e.g., $c_1, c_2, \ldots, c_k$). An assemblage with an ideal target-to-marker ratio for one target type (e.g., pollen, algae, terrestrial organic microfossils) is unlikely to be suitable for other potential target types in the same assemblage, given the narrow window of target-to-marker ratio suitability for the linear method (which, ideally, should be close to $\hat{u} = 1$). Hence, the chances of achieving satisfactory concentration precision for multiple targets using the linear method becomes vanishingly small (without reprocessing for each target population [see point 1 above] or extensive data collection effort [see point 2 above]).

For the reasons above, there will be many instances when either the targets or exotic markers are disproportionately rare. This will result in low precision, reflected by large total error values and confidence intervals. It is worth noting that even in extreme cases where the target-to-marker ratios are exceptionally high or low, the linear method can still inform a qualitative assessment even with large error values. A very high (or very low) target-to-marker ratio is indicative of a very high (or very low) absolute abundance of targets. In this case, the linear method may still provide an approximate minimum (or maximum) value for these abundances (e.g., [85]). However, most studies would require more precise estimates. With the linear method, collecting sufficient count data to improve these precisions may require reprocessing, or an inordinate data collection time, both of which may be prohibitive for routine study. The 'FOVS method' (see below) aims to improve precision and/or sampling effort for samples with suboptimal target-to-marker ratios.

## Field-of-view subsampling method ('FOVS method'): limitations

The resilience of the FOVS method to the target-to-marker ratio circumvents a key limitation of the linear method. As a result, very large, extrapolated data sets can be obtained using the FOVS method, while bolstering the number of rare specimens. Despite the compounded error that this extrapolation entails, our simulated and empirical case studies (see the results section) demonstrate generally improved precision for absolute abundance estimates and/or decreased data collection effort. Furthermore, the simulations show that the FOVS method yields a more accurate approximation of precision. However, some limitations should be considered before applying this technique:

1. Unsuitable for relative abundances. The FOVS method is designed for absolute abundance (e.g., concentration) estimates. By counting entire fields of view in a rapid sequence (during the full count phase), it extrapolates large data sets by bypassing the collection effort of individual identifications. In contrast, a primary strength of the linear method is the identification of all target specimen types. A byproduct of the latter approach is a robust quantitative data set of the indigenous specimen populations. Hence, the linear method has the potential to provide concentrations for only a small number of targets (those with near-optimal target-to-marker ratios) while providing accurate, concurrent collection of compositional data (or relative abundance data) for the organic constituents of a sediment or sedimentary rock. Such relative abundance data form a cornerstone of pollen [86] and palynofacies [87] analyses. However, depending on the research question, the time-consuming determination of relative population abundances may not be necessary if absolute abundances are the primary goal.

2. Requires visual contrast. In visual observation applications, the FOVS method would work best when there is sufficient visual contrast between exotic markers and the indigenous specimens in the assemblage (Fig 2B). This contrast enables rapid and accurate identification of markers during the full counts. Without this, thorough observation of each field of view is required to produce accurate results, thus increasing data collection time (specifically, modifying $\omega$). The most common exotic markers utilized in organic microfossil studies today, modern *Lycopodium* spores, have typically undergone acetolysis during preparation which darkens and discolours them [25] (R. Muscheler & Å. Wallin, pers. comm., 2023). This provides them with sufficient contrast in modern or Quaternary assemblages, but can render them more difficult to distinguish from indigenous grains of some deep-time assemblages that have undergone darkening via thermal maturation (e.g., [88–91]). If this is problematic, an alternative method of increasing contrast is to utilize fluorescence microscopy. Owing to the distinctive autofluorescence response of some exotic grains (e.g., *Lycopodium* spores; Fig 6), the contrast between markers and other grains (particularly in thermally matured assemblages) can be greatly enhanced, thus expediting accurate data collection.

3. Sensitive to heterogeneous spatial distribution. The FOVS method is susceptible to heterogeneity in specimen distribution across the study surface area. This stems from the primary difference between the linear and FOVS methods: the spatial variance of common specimen abundance during the calibration count phase. If the number of specimens per field of view is highly variable during these counts (i.e., high $s_3$), the sample's total error ($\sigma_{Fx}$) will increase correspondingly, resulting in decreased precision. To circumvent this issue, it is essential that the spatial ranges of both the calibration and full counts are near equivalent. In other words, the regions chosen for both the full and calibration-count fields of view should overlap or be located close to each other, thus ensuring that specimen densities for both counts will be roughly equivalent (Fig 2A). In the case of organic microfossil slides, we recommend avoiding the slide margins where densities tend to be lower than the medial areas, since these may result in significant edge effects (Fig 2A). Our preliminary Monte Carlo computer experiments have shown that the dispersed nature of the FOVS method counting (as opposed to the contiguous counting in the linear method) may help to compensate for this heterogeneity in specimen distribution. However, we do not yet have quantitative data on this effect and should be the focus of future efforts.

To summarise, the FOVS method is designed for absolute abundance estimates of one or more targets with higher precision and/or reduced data collection effort than the linear method. However, it is not optimised for collecting relative abundances of diverse specimen categories. Moreover, to maximise the accuracy and precision of the FOVS method, we recommend: 1, high visible contrast between markers and other specimens; and 2, study area ranges for both calibration and full counts that are equivalent and homogeneous.

## Which technique to use?

Regardless of the count type, the most precise estimates of concentration are achieved when the target specimens (e.g., terrestrial organic microfossils, $c_t$) and exotic markers (e.g., introduced *Lycopodium* spores) are close to equivalent. However, the FOVS method is particularly resilient to disproportionate ratios.

We have provided a test for determining the superior (highest precision and lowest effort) method for a given assemblage (Eqn 20). Highly precise estimates for the necessary parameters of this test—specifically, $\omega$, $\hat{u}$ and $\overline{Y}_{3x}$—will only be available after a substantial amount of data collection (e.g., after the calibration counts of the FOVS method; see 'field-of-view subsampling (FOVS) method: operation'). However, sufficient estimates of two of the parameters (the target-to-marker ratio, $\hat{u}$, and the target density per field of view, $\overline{Y}_{3x}$) will generally be obvious upon cursory inspection of an assemblage. The third parameter (transition-to-count ratio, $\omega$) is the time variable, and is dependent on the target types, observation methods, and observer habits, which are largely consistent across assemblages in a study; this factor can be estimated even before inspecting a given assemblage. Hence, 'ballpark' values of these parameters will typically be easy to obtain; with these values, we recommend utilising Eqn 20 (Fig 1, steps 1, 2)—which is available on our user-friendly 'absolute abundance calculator' (S1 Text)—upon first examination of each assemblage for determining the most efficient count method.

In most cases, the FOVS method is more efficient except for those with extremely low specimen densities (very low $\overline{Y}_{3x}$) and/or where markers and targets are near-equivalent ($\hat{u} \approx 1$). We recommend that in cases where target-to-marker ratios approach 1:1, the 'linear method' of counting should be preferred. This is because this approach also provides additional data that may be important to the researcher (e.g., relative population abundances), and does not require additional conditions such as high marker contrast, or consistent areas for both calibration and full counts. In all other cases, the FOVS method should provide superior concentration and precision estimates.

The primary considerations are summarised in Table 3, and we have provided a flowchart (Fig 1) to assist in determining the most appropriate count method. We have also supplied the simulation Matlab code with which the reader may experiment to inform their own methodology (S16 Text). Many of the practical parameters discussed herein have been integrated into a user-friendly web-based application (see S1 Text). This app has been designed to calculate key absolute abundance outputs and assist in choosing the most efficient method for each assemblage; see S3 Table for a list of these parameters.

| **Consideration** | **Linear method** | **Field-of-view subsampling (FOVS) method** |
|---|---|---|
| Precision sensitivity to target-to-marker ratio ($\hat{u}$) | High | Low to moderate |
| Optimal range of target-to-marker ratio ($\hat{u}$) values[1] | c. 0.33–3 | c. 0.05–20 |
| Accuracy of error estimates | For $\hat{u} \approx 1$ | For all tested values of $\hat{u}$ |
| Concentrations of multiple target populations ($c_1, c_2, \dots, c_k$)? | No[2] | Yes |
| Suitable for relative abundance data? | Yes | No |
| Precision sensitivity to specimen heterogeneity[3] | Low | Moderate to high |
| Sampling effort sensitivity to common specimen density ($\overline{Y}_3$) | Low | Moderate |

**Table 3**. Summary of applicability for the two count methods discussed herein. [1]While maximum count precision is at $\hat{u} = 1$ [55], the optimal efficiency of each method is at $\hat{u} \approx 2$ (with target counts of ≥500; [22]). [2]It is very unlikely that near-optimal target-to-marker ratios of more than one target population will co-occur in a given assemblage. [3]This factor has not been quantified in this study.

Lastly, we have observed from preliminary experiments with the Monte Carlo simulations that the relative efficiency of the methods is largely dependent on the homogeneity of the specimen distributions. We have not yet quantified these observations, but given that lab-based samples may be quite heterogeneous, we recommend that this important question should be investigated in future work.

## Conclusions

When describing the absolute abundance method first introduced in the 1960s (the 'linear method', herein), Maher [22] stated (p. 154): "considering the potential of the [absolute abundance] method, it is surprising it is not more widely used." We echo this sentiment and aim to encourage the wider use of absolute abundance estimation by providing a new technique: the field-of-view subsampling (FOVS) method. Through a combination of computer-simulated and empirical data (terrestrial organic microfossils from the Permian–Triassic strata of the Tasmania Basin, Australia), the FOVS method demonstrates greater efficiency than the linear method under most conditions. The variables that had the greatest impacts on the relative efficiency of the methods were: 1, the ratio of targets to marker specimens; 2, the spatial density of targets; and 3, and the relative duration of field-of-view translation vs specimen counting. The suitability of the FOVS method across a broader range of data sets stems from the potential of this approach to glean multiple, precise absolute abundance estimates even from assemblages that might not be optimised for absolute abundance counts. The FOVS method has the added benefit of demonstrating more accurate precision estimates than the linear method in almost all cases. The FOVS method may also provide greater precision and/or lower sampling effort (than the linear method) for samples that contain highly heterogeneous specimen distributions, but this remains to be tested. We hope that the versatility of the new count technique will not only encourage a broader adoption of absolute abundance estimation in the future, but facilitate the re-examination of legacy data sets (e.g., curated fossil collections) previously considered inappropriate for absolute abundance estimation. This study has provided a stepwise process for choosing the optimal method for each data set, aided by a user-friendly software interface and the source code for the computer simulations herein.

While initially developed and applied to organic microfossil assemblages, the FOVS method could, in principle, be readily applied to any count data that involves: 1, area-based sampling; and 2, readily identifiable markers of known quantity.

## Acknowledgements

Christopher Fielding, Tracy Frank, Steve McLoughlin and Vivi Vajda for their thought-provoking comments and discussions regarding the utilisation of these methods. We also thank Ricardo Vilela and Francisco Porto for their assistance with the fluorescence microscopy. Thanks to Sam Slater (Swedish Museum of Natural History) and Richard V. Tyson for their discussions on the potential applications of the methods outlined herein. Thanks to Richard Bailey (University of Adelaide) for helpful discussions regarding sampling statistics. This project was funded by a Research Centre in Applied Geosciences (Science Foundation Ireland) grant awarded to CM (#13/RC/2092_P2).

# Supporting information 1

## Software interface and additional methods

This document includes: 1, details of the software-based interface for calculating the output parameters in this paper; 2, all additional supporting equations, and details of the statistical corrections used; and 3, the caption for the supporting Information figure (S17 Fig), which is an illustrated summary of key simulation data (S5–S15 Tables).

### Software interface for calculating parameters of the linear and FOVS methods

We devised an alternative way for users to calculate the quantitative parameters described in this paper. The 'absolute abundance calculator' is designed as a user-friendly interface that provides all the practical outputs for these methods, including concentration, uncertainty and effort. Moreover, the calculator assists in determining the most efficient method (either linear and FOVS) for a given sample, and a 'counting assistant' aids in rapid and efficient data collection.

The calculator has been designed using Microsoft Excel's Visual Basic for Applications (VBA). It uses macros—predefined sets of instructions that can be executed automatically—that aid users with certain tasks, such as automatically determining variables (e.g., $\omega$, $s_3$), a timer, a counting assistant that allows faster counting in comparison to physical methods (e.g., clickers), and a way to export all variables and counts to spreadsheets.

This 'absolute abundance calculator' is open-source and hosted in a GitHub repository, where its VBA code can be read, and the application downloaded. Software updates will be made available there, and users are free to use GitHub functionalities as a collaborative platform to raise any issues or suggest code changes.

The application is contained inside a single macro-enabled Microsoft Excel worksheet (file format extension: xlsm). Users are required to first enable macros to run this tool. Additional instructions are present in the README.md file present in the GitHub repository. Readers can access the 'absolute abundance calculator' here: https://github.com/Palaeomays/FOVS_vs_linear_methods.git.

## Analysis of precision as a function of effort and the ratio of common and rare grains

The precision of both methods will improve with increased effort. In this section, we analyse this more closely to quantify how much the precision improves, and how $\hat{u}$—the ratio of target-to-marker specimens (usually, common-to-marker specimens)—affects this improvement.

### Linear method

For the linear method, increasing effort ($e_L$) results in increased precision (i.e., decreased error, $\sigma_L$). In practice, this translates into an increased length of our observational 'window' on the study area, resulting in greater quantities of targets and markers. We can obtain an estimate of this improvement by calculating the rate of change of $\sigma_L$ for changes in $e_L$. Mathematically, this is expressed as the derivative of $\sigma_L$ with respect to $e_L$ as follows:

$$\frac{\partial}{\partial e_L}\sigma_L = -50\frac{A(1+\hat{u})}{{e_L}^2\sqrt{T+(1+\hat{u})\frac{A}{e_L}}} \xrightarrow[e_L\to\infty]{} -50\frac{A(1+\hat{u})}{{e_L}^2\sqrt{T}}. \qquad \text{(S1)}$$

You will notice that the output values of S1 Eqn will be negative (since $A$ and $\hat{u}$ are both positive), which is expected since the error should reduce with increasing effort. Specifically, linear method error decreases with the reciprocal of the square of the effort. We can also see that the improvement to the precision depends on the target-to-marker ratio ($\hat{u}$) as well as the density of the target specimens across the study area (via the $\omega/\,\overline{Y}_{3x}$ term in $A$).

Lastly, we can use S1 Eqn to rederive a result by Regal & Cushing [1] mentioned in the main text; specifically, that the highest precision estimates of the concentration are achieved when $\hat{u} = 1$ (i.e., $x = n$). In other words, the question we are asking is how the error changes as $\hat{u}$ changes, which is expressed through the derivative:

$$\frac{\partial}{\partial \hat{u}}\sigma_L = 50\frac{\left(\omega/\overline{Y}_{3x}\right) + 1 - \hat{u}^{-2}}{e_L\sqrt{T + (1+\hat{u})\dfrac{A}{e_L}}} \geq 0. \tag{S2}$$

We see that the output values are always non-negative, since both $\omega$ and $\overline{Y}_{3x}$ are greater than zero, and $\hat{u} \geq 1$. (Recall that in this formalism, we have assumed $x \geq n$.) So, this means that the error ($\sigma_L$) increases (=precision decreases) as the target-to-marker ratio ($\hat{u}$) increases, with $\hat{u} = 1$ giving the minimum error, exactly as Regal and Cushing [1] claimed.

**FOVS method**

To validly compare the efficiencies of the FOVS and linear methods, we need to characterise the relationship between FOVS method error ($\sigma_F$) and effort ($e_F$). As noted in the main text, a first step is to determine the optimal ratio of calibration and full-count fields of view, which will simplify the relationship to a single dimension. This was done by taking the derivative of the FOVS method error function (Eqn 5) and setting it equal to zero, resulting in Eqns 14 and 15. With these optimal values, we can then derive expressions analogous to S1 and S2 Eqns. These derivations are detailed below.

By differentiating Eqn 11 with respect to each of the two independent variables ($N_{3C}$ and $N_{3F}$), we can calculate the effect that the target-to-marker ratio ($\hat{u}$) and target specimen density ($\overline{Y}_{3x}$) have on FOVS method precision with the following two equations:

$$\frac{\partial}{\partial N_{3C}}\sigma_F(N_{3C},N_{3F}) = -50\frac{1}{N_{3C}{}^2\overline{Y}_{3x}\sqrt{T+\frac{1}{N_{3C}\overline{Y}_{3x}}+\frac{\hat{u}}{N_{3F}\overline{Y}_{3x}}}} \tag{S3}$$

and

$$\frac{\partial}{\partial N_{3F}}\sigma_F(N_{3C},N_{3F}) = -50\frac{\hat{u}}{N_{3F}{}^2\overline{Y}_{3x}\sqrt{T+\frac{1}{N_{3C}\overline{Y}_{3x}}+\frac{\hat{u}}{N_{3F}\overline{Y}_{3x}}}}\,. \tag{S4}$$

Since $N_{3F}$ and $N_{3C}$ determine the effort for the FOVS method, these equations suggest that the increase in precision decreases with the reciprocal of the square of the effort, akin to the linear method (S1 Eqn). This should follow because of the linear relationship between $N_{3C}$, $N_{3F}$ and $e_F$ in Eqn 7.

To more explicitly characterise how error changes for increased effort, we use Eqn 17 (which used the optimal numbers of calibration and full count fields of view to express error as a function of effort) and differentiate it with respect to $e_F$ to find

$$\frac{\partial}{\partial e_F}\sigma_F = -\frac{50\left([1+\hat{u}]\omega+2\overline{Y}_{3x}+2\sqrt{[\omega+\overline{Y}_{3x}][\omega\hat{u}+\overline{Y}_{3x}]}\right)}{e_F^2\overline{Y}_{3x}\sqrt{T+\frac{1}{e_F\overline{Y}_{3x}}\left([1+\hat{u}]\omega+2\overline{Y}_{3x}+2\sqrt{[\omega+\overline{Y}_{3x}][\omega\hat{u}+\overline{Y}_{3x}]}\right)}}\xrightarrow[e_F\to\infty]{}-\frac{50\left([1+\hat{u}]\omega+2\overline{Y}_{3x}+2\sqrt{[\omega+\overline{Y}_{3x}][\omega\hat{u}+\overline{Y}_{3x}]}\right)}{e_F^2\overline{Y}_{3x}\sqrt{T}}\,. \tag{S5}$$

This shows that, similar to S1 Eqn, as we increase the amount of effort, the improvement in precision decreases as the reciprocal of the square of the effort. So, we can conclude that the FOVS method has similar asymptotic behaviour to the linear method, and so the FOVS method is no worse that the linear method.

Lastly, we can also find the value of $\hat{u}$ that minimises the error by calculating an expression analogous to S2 Eqn, which was formulated for the linear method. We differentiate S5 Eqn with respect to $\hat{u}$ to obtain

$$\frac{\partial}{\partial \hat{u}}\sigma_F = \frac{50\omega\left(\omega\hat{u} + \overline{Y}_{3x} + \sqrt{[\omega + \overline{Y}_{3x}][\omega\hat{u} + \overline{Y}_{3x}]}\right)}{e_F\overline{Y}_{3x}\left(\omega\hat{u} + \overline{Y}_{3x}\right)\sqrt{T + \frac{1}{e_F\overline{Y}_{3x}}\left([1 + \hat{u}]\omega + 2\overline{Y}_{3x} + 2\sqrt{[\omega + \overline{Y}_{3x}][\omega\hat{u} + \overline{Y}_{3x}]}\right)}} \geq 0. \qquad (S7)$$

From this equation, we see that any increase in $\hat{u}$ will increase the error. Therefore, the maximum precision is obtained where $\hat{u} = 1$, as it was for the linear method.

## Field-of-view subsampling (FOVS) method variant: if markers ($n$) are more common than targets ($x$)

The 'common' and 'rare' specimen types may differ between assemblages, as a function of concentration and/or number of introduced exotic markers. As such, the type of specimen for the calibration counts may vary between assemblages, and—for the purposes of precise concentration estimates—the most common specimen group should be preferred. This is because a larger sample size should result in lower standard deviations for the calibration counts. However, for most applications of this method, the targets ($x$) would be preferred as the more common specimen type, since disproportionately high counts of markers ($n$) do not provide additional details on the population of interest for our research (e.g., the indigenous fossils of an assemblage, as in the empirical case study presented herein). Put another way: our time is better spent collecting target, rather than marker, data, since these provide additional information about the assemblage beyond the target concentrations (e.g., relative abundances of the targets to other populations, occurrence of biostratigraphic index taxa, fossil preservation quality, etc.).

When the markers are the subject of the calibration counts, then their extrapolated sample abundance ($\hat{n}$) will need to be estimated from their calibration-count mean ($\overline{Y}_{3n}$) multiplied by the total number of full-count fields of vi$\hat{n}$ew ($N_{3F}$). This is analogous to Eqn 3, but $\hat{n}$ is substituted for $\hat{x}$ so that

$$\hat{n} = \overline{Y}_{3n} \times N_{3F}. \tag{S8}$$

In this variant of the method, whereby markers are the subjects of the calibration counts, the concentrations of organic microfossils ($c_{Fn}$) can be estimated with a modified version of Eqn 1. Specifically, we can substitute $n$ with $\hat{n}$, and use the total number of target specimens from the full counts ($x$) as follows:

$$c_{Fn} = \frac{x \times N_1 \times \overline{Y}_1}{\hat{n} \times \overline{V}} \tag{S9}$$

Similarly, the calculation of total error (Eqn 5) necessitates a slight modification when the markers are the subject of the calibration count:

$$\sigma_{Fn} = 100 \times \sqrt{\left(\frac{s_{1P}}{\sqrt{N_1}}\right)^2 + \left(\frac{\sqrt{x}}{x}\right)^2 + \left(\frac{s_{3P}}{\sqrt{N_{3C}}}\right)^2}. \tag{S10}$$

As discussed for Eqn 11, for the purposes of the mathematical analysis, we assume the Poisson approximation $\left(\frac{n}{\sqrt{n}}\right)^2$ for the sample standard deviation $\left(\frac{s_{3P}}{\sqrt{N_{3C}}}\right)^2$. Here, the number of markers is given by $n = \overline{Y}_{3n} N_{3C}$ and the number of targets is $x = \overline{Y}_{3x} N_{3F}$. Then, by the same reasoning as Eqn 8, the following approximations can be made: $x = n \times \hat{u}$ for the linear method, and $\overline{Y}_{3x} = \overline{Y}_{3n} \times \hat{u}$ for the FOVS method. So, in cases where markers are more common than targets, we substitute these values into several of the equations expressed in the manuscript. Of particular importance to discriminating between the FOVS vs linear method is Eqn 16, which becomes

$$\delta^*(Y_{3n}) = \frac{N_{3F}^*(e_F)}{N_{3C}^*(e_F)} = \frac{1}{\hat{u}}\sqrt{\frac{\omega + Y_{3n}}{\omega\hat{u}^{-1} + Y_{3n}}}\ . \qquad \text{(S11)}$$

Hence, the equation for $\overline{Y}_{3n}^*$ (modified from Eqn 20), which indicates the critical density of markers (not targets) per field of view, can be calculated

$$\overline{Y}_{3n}^* = 2\omega\frac{\hat{u}^2 + \sqrt{\hat{u}^3(1+\hat{u}[\hat{u}-1])}}{\hat{u}(\hat{u}+1)(\hat{u}-1)^2}, \qquad \text{(S12)}$$

whereby, if $\overline{Y}_{3n} > \overline{Y}_{3n}^*$, then the FOVS method should be used.

By having the marker specimens as the subject of the calibration counts, the two-dimensional problem of FOVS method error outlined in Eqn 11 becomes

$$\sigma_F(N_{3C}, N_{3F}) = 100\sqrt{T + \frac{1}{N_{3\mathrm{F}}(\overline{Y}_{3n} \times \hat{u})} + \frac{1}{N_{3C}\overline{Y}_{3n}}}. \qquad \text{(S13)}$$

When the markers are more common than targets, the optimal counts of calibration fields of view ($N_{3C}^*$) and full count fields of view ($N_{3F}^*$) in the FOVS method are given by

$$N_{3C}^*(\bar{\sigma}) = \frac{1}{(\bar{\sigma}/100)^2 - T}\left(\frac{\sqrt{\left[\overline{Y}_{3n} + \omega\right]} + \sqrt{\left[\overline{Y}_{3n} + \frac{\omega}{\hat{u}}\right]}}{\overline{Y}_{3n}\sqrt{\left[\overline{Y}_{3n} + \omega\right]}}\right) \qquad \text{(S14)}$$

and

$$N_{3F}^*(\bar{\sigma}) = \frac{\frac{1}{\hat{u}}}{(\bar{\sigma}/100)^2 - T}\left(\frac{\sqrt{\left[\overline{Y}_{3n} + \omega\right]} + \sqrt{\left[\overline{Y}_{3n} + \frac{\omega}{\hat{u}}\right]}}{\overline{Y}_{3n}\sqrt{\left[\overline{Y}_{3n} + \frac{\omega}{\hat{u}}\right]}}\right). \qquad \text{(S15)}$$

Note: for this study, we assume that the targets are more common than the exotic markers, unless stated otherwise.

## Effort standardisation for simulation precision estimates

The purpose of the simulation was to compare the linear and FOVS method precisions for an equivalent amount of effort. However, the estimated effort of each method is a non-trivial combination of various deterministic and random factors (Eqns 6 and 7). Hence, it was not possible to set up the simulation to produce identical collection effort estimates for both methods. However, with some trial-and-error, we found sets of parameters that yielded roughly equal efforts, which we used as the inputs for the simulation (see simulation output data in S5–S15 Tables).

But with unequal effort, the reliability of any comparison between precisions of the two methods will suffer. To correct for this, we used rescaled total error estimates for each method ($\tilde{\sigma}_L$ for the linear method; $\tilde{\sigma}_F$ for the FOVS method)

$$\tilde{\sigma}_L = \sigma_L \frac{e_L}{\overline{e}} \tag{S16}$$

$$\tilde{\sigma}_F = \sigma_F \frac{e_F}{\overline{e}} \tag{S17}$$

where $\overline{e}$ is the average effort $(e_L + e_F)/2$. This rescaling will increase the error of the method that is associated with more work, while decreasing the error of the method associated with less work, making for a fairer comparison. Row 3 of S5–S15 Tables contain these values, and they are plotted in S17 Fig.

## Deviations from exact abundances

As discussed in "case study: computer simulation", we know the exact numbers of specimens in the simulated data sets in each virtual study area, hence: the exact concentration ($c_{exact}$). This allows us to compare the standard deviation of the concentration prediction of the two methods ($c_L$ and $c_F$) from this known true value ($c_{exact}$). Using the conventional sampling standard deviation, the equation for this is

$$\sigma_{exact,M} = 100\sqrt{\frac{1}{(N-1)\times c_4(N)^2}\sum_{j=1}^{N}\left(\frac{c_M - c_{exact}}{c_{exact}}\right)^2}, \tag{S18}$$

where $N$ is the number of iterations of the simulation, the subscript $M$ denotes the chosen method for the calculation (linear $[L]$ or FOVS $[F]$ method), and $c_4(N)$ is the standard deviation bias correction (see S2 Table and '$c_4$ correction' below). Since the error calculations Eqns 2 and 5 are empirical approximations of the quantity in S18 Eqn, we can use the simulations to gauge the accuracy of each method to predict the error.

To keep the comparison between the methods fair, we need to apply the effort standardisations as defined in S16 and S17 Eqns, giving us the following two equations:

$$\tilde{\sigma}_{exact,L} = \sigma_{exact,L}\frac{e_L}{\overline{e}} \tag{S19}$$

$$\tilde{\sigma}_{exact,F} = \sigma_{exact,F}\frac{e_F}{\overline{e}}. \tag{S20}$$

The results are contained on rows 4, 5 and 6 of S5–S15 Tables.

## Statistical corrections for data sets

### Finite population correction

Note that Eqn S18 is defined explicitly for the known finite number of target specimens on the virtual study area. As mentioned in the main text (see 'case study: computer simulation'), there are finite population effects that can become important when sampling too much area of any finite population (e.g., virtual study areas in our simulations, or microfossil slides). In particular, the standard deviations of samples from a finite population will be lower than those from an infinite population, and this discrepancy inflates as the sample size increases. In order to ensure a correct comparison between the exact (effort-scaled) errors in S19 and S20 Eqns and the predicted errors in Eqns 2 and 5, we also calculate finite-population versions of $\sigma_L$ and $\sigma_F$ (see [2], p. 83) in S21 and S22 Eqns, respectively:

$$\hat{\sigma}_L = \frac{100\ e_L}{\bar{e}} \times \sqrt{\left(\frac{\sqrt{x}}{x}\right)^2 \left(\frac{\bar{\bar{x}} - x}{\bar{\bar{x}}}\right) + \left(\frac{\sqrt{n}}{n}\right)^2 \left(\frac{\bar{\bar{n}} - n}{\bar{\bar{n}}}\right)} \tag{S21}$$

and

$$\hat{\sigma}_F = \frac{100\ e_F}{\bar{e}} \times \sqrt{\left(\frac{s_{3P}}{\sqrt{N_3}}\right)^2 \left(\frac{\bar{\bar{x}} - x}{\bar{\bar{x}}}\right) + \left(\frac{\sqrt{n}}{n}\right)^2 \left(\frac{\bar{\bar{n}} - n}{\bar{\bar{n}}}\right)}. \tag{S22}$$

Note that these equations have the expected behaviour: as $x$ and $n$ approach the full population counts ($\bar{\bar{x}}$ and $\bar{\bar{n}}$, respectively), the errors approach zero; while for small $x$ and $n$ counts, the errors approximate the infinite-population cases in Eqns 2 and 5.

Comparisons of the scaled precision estimates between the two methods are illustrated in S17 Fig. We further note the finite population correction would not be typically required for working with real-world microfossil data, if the proportion of the slide covered by the total field-of-view areas is small. Moreover, the number of fossils per slide is itself a random variable, which offsets the reduction in variance as the total field-of-view area increases.

### $c_4$ correction

As mentioned in the main text, we have applied the $c_4$ correction to the standard deviation calculations of $s_{3P}$ in Eqn 5, and this was done for both the simulated and real fossil data. To calculate the correction, we use the expression in S2 Table.

The correction is also used in S18 Eqn; however, the $N$ in that equation refers to the number of iterations of the Monte Carlo simulation that are run to generate the data for averaging. Our simulations were run 1,000,000 times (i.e., $N = 10^6$); under these conditions, the biasing correction in S18 Eqn is $c_4(N) \approx 0.99999975$, which we approximate with $c_4(N) = 1$. (In the code of the simulation, we have explicitly included the approximation $c_4(N) = 1$ for $N > 341$ to avoid numerical errors, as the calculation software used herein (Matlab) provided infinite answers for $N \geq 344$.) The problem lies in the calculations of the gamma functions ($\Gamma$; see S2 Table), which are equivalent to factorial functions. These functions grow extremely fast as $N$ increases, and very quickly exceeds the calculation precision of standard computer packages. By carefully adjusting the precision (or using software dedicated to this process) the accuracy can be increased; however, to within the accuracy reported within this study, we did not need a better approximation for S18 Eqn.

**Jensen's inequality**

When calculating the concentration mean across multiple iterations of the same sample, an additional statistical correction needed to be accounted for: Jensen's inequality [3] (ch. 8.3). This implies that $\mathbb{E}(1/X) \geq 1/\mathbb{E}(X)$, where $X$ is a random variable with a non-zero mean. In other words, the average of a reciprocal is always greater than or equal to the reciprocal of the average. Since the calculations of the concentrations include a random variable in the denominator ($n$) for both the linear (Eqn 1) and FOVS (Eqn 4) methods, Jensen's inequality means that if we take the average of the concentration estimate from each simulation, then we overestimate the true average concentration. If results from multiple samples are being averaged over, then Jensen's inequality will need to be accounted for, regardless of the type of data (simulated or empirical). Since all empirical (microfossil) data analysed herein are from single samples, this correction was not needed.

There are several ways to correct for this [4], the choice of which depends on the absolute values of counted specimens (e.g., if our sample happened to have no specimens counted in the sampling region, we would be dividing by zero). For the purposes of our simulations, since our parameters never resulted in zero exotic marker counts, we could use the direct method of calculating the reciprocal average $N_{3F}/n$ (i.e., the number of full-count fields of view, which was fixed for each iteration of the simulations, divided by the total number of markers counted) as an unbiased estimator, since this is exactly $1/\mathbb{E}(X)$. However, if the reader's experimental setup is such that zero exotic marker counts are possible, then other estimators of the reciprocal of binomial proportions may be of use; see [4].

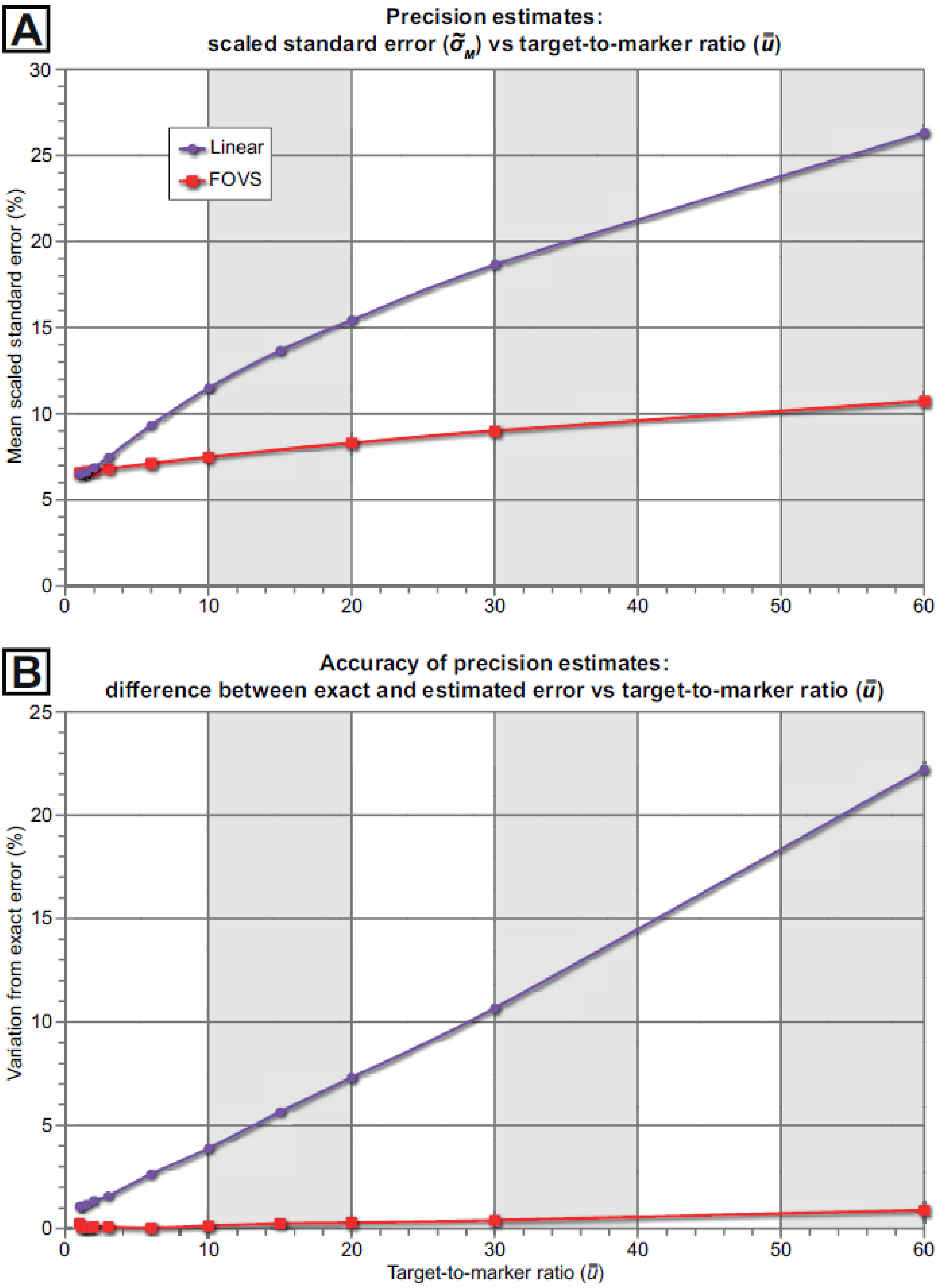

**S17 Fig. Simulated precisions of the linear and FOVS methods for different target-to-marker ratios ($\bar{\bar{u}}$). A**, Total standard error scaled to standardised effort ($\tilde{\sigma}_M$), expressed as %;

see S16 and S17 Eqns. **B**, Percentage difference between exact total standard error ($\tilde{\sigma}_{exact,M}$) and estimated total standard error (including the finite population correction; FPC), expressed as %; i.e., the values of S21 Eqn relative to S19 Eqn (linear method) or S22 Eqn relative to S20 Eqn (FOVS method). See S5–S15 Tables for the data expressed here.

# Supporting information 2

**S2 Table. List of statistical terms and their descriptions.** Confidence interval functions follow Maher [1], as updated by Mertens *et al*. [2]. †Note: In this paper we have used the factor $1/(N-1)$ ('Bessel's correction'; [3]) to give an unbiased estimator of the sample variances and sample standard deviations.

| Term | Description | Method(s) |
|---|---|---|
| PRIMARY INPUTS AND OUTPUTS FOR CONCENTATION ESTIMATES | | |
| $\boldsymbol{c}$ **(variants:** $\boldsymbol{c_L}$, $\boldsymbol{c_{Fx}}$, $\boldsymbol{c_{Fn}}$**)** | Concentration of target specimens per unit mass (or volume); see Eqns 1 or 4. Unless specified, this is derived by the 'linear method' (i.e., $c = c_L$). $c_F$ indicates concentrations calculated from the FOVS method; $c_{Fx}$ or $c_{Fn}$ are used when the target or marker specimens are the foci of the calibration, respectively. In the microfossil case study herein, $c$ is the concentration of organic microfossils per gram of dried sediment (grains/g). | Both |
| $\boldsymbol{c_t}$ | Concentration of terrestrial organic microfossils per gram of dried sediment (grains/g). This is used when the total terrestrial microfossil count is designated the target population of a microfossil assemblage. | Both |
| $\boldsymbol{x}$ **(variant:** $\boldsymbol{\bar{\bar{x}}}$**)** | Number of counted target specimens in a sample, while the total number of targets in a study area is denoted $\bar{\bar{x}}$. For the FOVS method, the target specimens of the calibration counts are typically the most common specimen type. If the markers are more common, see S1 Text. | Both |
| $\boldsymbol{n}$ **(variant:** $\boldsymbol{\bar{\bar{n}}}$**)** | Number of counted exotic markers (e.g., *Lycopodium* spores) in a sample, while the total number of markers in a study area is denoted $\bar{\bar{n}}$. | Both |
| $\boldsymbol{\hat{u}}$ **(variant:** $\boldsymbol{\bar{\bar{u}}}$**)** | Estimate of the target-to-marker ratio in a sample count. If all targets and markers in the population were counted, their ratio would provide the true population target-to-marker value in a study area ($\bar{\bar{u}}$, where $\bar{\bar{u}} = \frac{\bar{\bar{x}}}{\bar{\bar{n}}}$). However, this is impractical for routine work. So, samples of targets and markers provide estimates of $\bar{\bar{u}}$ based on the ratio ($\hat{u}$) in a given sample count ($\hat{u} = x/n$) for the linear method, and $\hat{u} = \overline{Y}_{3x}/\overline{Y}_{3n}$ for the FOVS method. | Both |
| $\boldsymbol{N_1}$ | Number of doses of exotic marker specimens. In the microfossil case study herein, these doses are tablets of *Lycopodium* spores. The details of the *Lycopodium* tablets utilised in the microfossil case study were provided by Lund | Both |
| | University (see 'case study: Permian–Triassic organic microfossils of southeastern Australia'). | |
| $\overline{Y}_1$ | Mean number of exotic markers for one dose (e.g., number of *Lycopodium* spores in one tablet). | Both |
| $s_1$ | Sample standard deviation† for one dose of exotic markers (e.g., standard deviation of *Lycopodium* spores in one tablet). | Both |
| $s_{1P}$ | Proportional sample standard deviation† of the number of exotic markers per dose $\left(s_{1P} = \frac{s_1}{\overline{Y}_1}\right)$. | Both |
| $s_m$ | Standard deviation of exotic markers added to the sample ($s_m = \sqrt{N_1} \times s_1$). | Both |
| $m$ | Total number of exotic markers added to the sample. | Both |
| $\overline{m}$ | Estimated number of exotic markers added to the sample ($\overline{m} = N_1 \times \overline{Y}_1$). | Both |
| $T$ | Error contribution from the exotic marker doses, e.g., *Lycopodium* tablets ($T = {s_{1P}}^2/N_1$). | Both |
| $N_2$ | Total number of samples combined for the concentration estimate. | Both |
| $\overline{Y}_2$ | Mean sample mass (or volume); for single samples, this is the specific sample mass (or volume). | Both |
| $\overline{V}$ | Total mass (or volume) of samples ($\overline{V} = N_2 \times \overline{Y}_2$). | Both |
| $s_2$ | Standard deviation of sample mass (or volume); for single samples, this can be approximated by the square root of the mass (or volume). Hence, if $N_2 = 1$, then $s_2 = \sqrt{\overline{Y}_2}$. | Both |
| $s_V$ | Standard deviation of mass (or volume) in sample ($s_V = \sqrt{N_2} \times s_2$). | Both |
| $N_{3C}$ | Number of fields of view counted during the calibration counts. | FOVS |
| $N_{3F}$ | Number of fields of view counted during the full counts. | FOVS |
| $\overline{Y}_3$ | Mean specimens (typically the target specimens, $x$) in each field of view. This is a measure of specimen density for the total sample area. If used as an estimate of target specimens per field of view, then $\overline{Y}_3 = \overline{Y}_{3x}$ (see Eqn 3); if used for markers, then $\overline{Y}_3 = \overline{Y}_{3n}$ (see S8 Eqn). | FOVS |
| $s_3$ | Sample standard deviation† for the common specimens (typically $x$, although this might be substituted for $n$; see S10 Eqn) counted during the calibration counts. | FOVS |
| $c_4$ | Correction factor to achieve an unbiased estimator of the population standard deviation. This factor is particularly important for small sample sizes, where the bias on the sample standard deviation can result in major differences from the population standard deviation. (See $\hat{s}_3$ below for the application of the $c_4$ factor.) | FOVS |
| $\hat{s}_3$ | Estimator of the population standard deviation† for the common specimens (typically $x$, although this might be | FOVS |
| | substituted for $n$ in cases where the latter is more common; see S10 Eqn) counted during the calibration counts. Following Gurland & Tripathi [4], this unbiased estimator is calculated:<br><br>$\hat{s}_3 = \frac{s_3}{c_4(N_{3C})}$, where<br><br>$c_4(N_{3C}) = \sqrt{\frac{2}{N_{3C}-1} \times \frac{\Gamma\left(\frac{N_{3C}}{2}\right)}{\Gamma\left(\frac{N_{3C}-1}{2}\right)}}$.<br><br>In the above formulation, $\Gamma$ is the 'gamma function' [5]. | |
| $\boldsymbol{s_{3P}}$ | Proportional sample standard deviation† of the number of common specimens in the calibration counts $\left(s_{3P} = \frac{\hat{s}_3}{\overline{Y}_3}\right)$. | FOVS |
| $\boldsymbol{\omega}$ | Field-of-view transition effort factor, equal to the mean transition time between fields of view divided by the mean count time for each specimen. | Both |
| $\boldsymbol{\hat{x}}$ | Extrapolated number of counted target specimens for the full counts $\left(\hat{x} = \overline{Y}_{3x} \times N_{3F}\right)$; see formula in Eqn 3. | FOVS |
| $\boldsymbol{\hat{n}}$ | Extrapolated number of counted marker specimens for the full counts $\left(\hat{n} = \overline{Y}_{3n} \times N_{3F}\right)$; see formula in S8 Eqn. | FOVS (variant) |
| $\boldsymbol{A}$ | The degree of effort for each target specimen $\left(\frac{\omega}{\overline{Y}_3} + 1 + \frac{1}{\hat{u}}\right)$; see formula in Eqn 9. | Linear |
| $\boldsymbol{N_{3C}^*}$ | Optimal number of calibration-count fields of view; see formulae in Eqns 14 and 23 (or variant S14 Eqn, which we recommend if $n > x$). | FOVS |
| $\boldsymbol{N_{3F}^*}$ | Optimal number of full-count fields of view; see formulae in Eqns 15 and 24 (or variant S15 Eqn, which we recommend if $n > x$). | FOVS |
| $\boldsymbol{\delta^*}$ | The optimal field-of-view count ratio ($\delta^* = \frac{N_{3F}^*}{N_{3C}^*}$); see formula in Eqn 16 (or variant S11 Eqn, which we recommend if $n > x$). | FOVS |
| $\boldsymbol{\overline{Y}_3^*}$ | Critical value of field-of-view target density [$\overline{Y}_3$] whereby either the FOVS or linear method is the superior choice. This parameter is utilised for the 'method determination test'. If the most common specimens are the targets, then $\overline{Y}_3^* = \overline{Y}_{3x}^*$ (see Eqn 20); if the most common specimens are markers, then $\overline{Y}_3^* = \overline{Y}_{3n}^*$ (see variant S12 Eqn, which we recommend if $n > x$). | Both |
| $\boldsymbol{\sigma_L}$ | Total standard error on the concentration (in %); see formulae in Eqns 2 and 10. | Linear |
| $\boldsymbol{\sigma_F}$ | Total standard error on the concentration (in %). If the most common specimens are the targets, then $\sigma_F = \sigma_{Fx}$ (see Eqns 5 and 11); if the most common specimens are markers, then $\sigma_F = \sigma_{Fn}$ (see S10 Eqn). | FOVS |
| $\overline{\sigma}$ | User-defined desired level of error (in %); see formulae in Eqns 21 and 25. | Both |
| $e_L$ | Data collection effort in dimensionless units of work; see formulae in Eqns 6 and 9. | Linear |
| $e_F$ | Data collection effort in dimensionless units of work; see formula in Eqn 7. | FOVS |
| $e_L(\overline{\sigma})$ | Data collection effort for a user-defined error ($\overline{\sigma}$), as a function of effort; see formula in Eqn 21. | Linear |
| $e_F(\overline{\sigma})$ | Data collection effort for a user-defined error ($\overline{\sigma}$), as a function of effort; see formula in Eqn 25. | FOVS |
| CONFIDENCE INTERVAL FUNCTIONS | | |
| $\alpha$ | $= \tan^{-1}[(\overline{m}/s_m)\,/(\overline{V}/s_V)]$ | Linear |
| $\beta$ | $= \sin^{-1}[1\,/\sqrt{(\overline{m}/s_m)^2 + (\overline{V}/s_V)^2}\,]$ | Linear |
| $(m/V)_{\max}$ | $= [s_m \times \tan(\alpha + \beta)]\,/\,s_V$ | Linear |
| $(m/V)_{\min}$ | $= [s_m \times \tan(\alpha - \beta)]\,/\,s_V$ | Linear |
| $\hat{u}_{max}$ | $= \dfrac{\hat{u} + \left[\frac{1}{(2n)}\right] + \sqrt{\left[\frac{\hat{u}(1+\hat{u})}{n}\right] + \left[\frac{1}{4n^2}\right]}}{1 - \left(\frac{1}{n}\right)}$ | Linear |
| $\hat{u}_{min}$ | $= \dfrac{\hat{u} + \left[\frac{1}{(2n)}\right] - \sqrt{\left[\frac{\hat{u}(1+\hat{u})}{n}\right] + \left[\frac{1}{4n^2}\right]}}{1 - \left(\frac{1}{n}\right)}$ | Linear |
| $\log \hat{u}$ | $= \log(x/n)$ | Linear |
| $s_{\log\hat{u}}$ | $= (\log \hat{u}_{\max} - \log \hat{u}_{\min})/2$ | Linear |
| $\overline{\log m/V}$ | $= \log(\overline{m}/\overline{V})$ | Linear |
| $s_{\log(m/V)}$ | $= [\log(m/V)_{\max} - \log(m/V)_{\min}]/2$ | Linear |
| $\log F$ | $= Z\sqrt{(s_{\log\hat{u}})^2 + (s_{\log(m/V)})^2}$ | Linear |
| Log limit | $= (\log \hat{u} + \overline{\log m/V}) \pm \log F$ | Linear |
| $Z$-score | This value denotes the distance (in standard deviation units) of an observed value from the mean. Below are some example $Z$-scores for commonly used confidence levels.<br>90% confidence level: $Z \sim 1.65$<br>95% confidence level: $Z \sim 1.96$<br>99% confidence level: $Z \sim 2.58$… | Linear |
| $CI_{max}$ | Confidence interval maximum, $CI_{max} = \frac{\hat{u} \times \overline{m} \times F}{\overline{V}}$ | Linear |
| $CI_{min}$ | Confidence interval minimum, $CI_{min} = \frac{\hat{u} \times \overline{m}}{\overline{V} \times F}$ | Linear |

# Supporting information 3

**S3 Table. Key input and output parameters for the absolute abundance calculations.**

Approximations for the 'pre-collection' outputs can all be achieved before completing data collection for a given sample (e.g., prior to, or during, the calibration counts of the FOVS method), and can guide a user's choice of method and/or data collection parameters. # Confidence interval functions provided by Maher [1] as updated by Mertens et al. [2], terms listed in S2 Table.

| **Function** | **Method** | **Input parameters** | **Output parameter** | **Equation** |
|---|---|---|---|---|
| PRE-COLLECTION OUTPUTS | | | | |
| **1) Optimal field-of-view count ratio** | FOVS | $\hat{u}, \overline{Y}_3, \omega$ | $\delta^*$ | Eqn 16 (or S11) |
| **2) Method determination 1 (minimum specimen density per field of view for FOVS method superiority)** | Both | $\hat{u}, \overline{Y}_3, \omega$ | $\overline{Y}_3^*$ | Eqn 20 (or S12) |
| **3) Optimal number of calibration-count fields of view (for user-defined desired error)** | FOVS | $N_1, \overline{Y}_1, s_1, \hat{u}, \overline{Y}_3, \omega, \bar{\sigma}$ | $N_{3C}^*$ | Eqn 14 (or S14) |
| **4) Optimal number of full-count fields of view (for user-defined desired error)** | FOVS | $N_1, \overline{Y}_1, s_1, \hat{u}, \overline{Y}_3, \omega, \bar{\sigma}$ | $N_{3F}^*$ | Eqn 15 (or S15) |
| **5) Predicted data collection effort (for user-defined desired error)** | Linear | $N_1, \overline{Y}_1, s_1, \hat{u}, \overline{Y}_3, \omega, \bar{\sigma}$ | $e_L(\bar{\sigma})$ | Eqn 21 |
| **6) Predicted data collection effort (for user-defined desired error)** | FOVS | $N_1, \overline{Y}_1, s_1, \hat{u}, \overline{Y}_3, \omega, \bar{\sigma}$ | $e_F(\bar{\sigma})$ | Eqn 25 |
| POST-COLLECTION OUTPUTS | | | | |
| **7) Concentration** | Linear | $N_1, \overline{V}, \overline{Y}_1, x, n$ | $c_L$ | Eqn 1 |
| **8) Total error** | Linear | $N_1, \overline{Y}_1, x, n, s_1$ | $\sigma_L$ | Eqn 2 |
| **9) Confidence intervals** | Linear | $N_1, \overline{Y}_1, x, n, s_1, N_2, s_2, Z$ | $CI_{min}, CI_{max}$ | # |
| **10) Concentration** | FOVS | $N_1, \overline{V}, \overline{Y}_1, x, n, \overline{Y}_3, f$ | $c_{Fx}$ | Eqn 4 (or S9) |
| **11) Total error** | FOVS | $N_1, \overline{Y}_1, s_1, x, n, \overline{Y}_3, s_3$ | $\sigma_{Fx}$ | Eqn 5 (or S10) |

# Supporting information 5–15

## Simulation summary data tables

The following tables summarise the simulated data sets. Instructions for these simulations are provided in S16 Text, and the data are illustrated in S17 Fig.

**S5 Table. Simulation summary data (simulation conditions 1 of 11).** Comparison table of terrestrial organic microfossil concentration estimates ($c_M$) from Eqn 1 (when $M = L$) or Eqn 4 (when $M = F$), and their associated errors and sampling efforts from a simulated data set of randomly distributed target and exotic specimens.

Parameters: total targets in assemblage $= 30{,}000$; total markers in assemblage $= 30{,}000$; target-to-marker ratio = 1:1 (i.e., $\bar{\bar{u}} = 1$); $x$ count (linear method) $= 482$; simulated iterations $= 10^6$; $\omega = 2$; $N_{3C} = 17$; $N_{3F} = 17$; $\overline{Y}_3 = 27$; $\overline{Y}_3^* = \infty$.

Since $\overline{Y}_3 < \overline{Y}_3^*$, the linear method is more efficient for this assemblage.

| Parameter estimates | Linear method ($M = L$) | FOVS method ($M = F$) |
|---|---|---|
| **1) Concentration ($c_M$; specimens/unit size), mean** | 30004 | 30000 |
| **2) Sampling effort ($e_M$; time units), mean** | 999.7 | 986.0 |
| **3) Estimated scaled standard error ($\widetilde{\sigma}_M$; %), mean (S16 and S17 Eqns)** | 6.492 | 6.587 |
| **4) Exact total standard error from true concentration ($\widetilde{\sigma}_{exact,M}$; %) (S19 and S20 Eqns)** | 6.510 | 6.553 |
| **5) Estimated scaled standard error, with finite population correction ($\widehat{\sigma}_M$; %), mean (S21 and S22 Eqns)** | 6.440 | 6.537 |
| **6) Difference between rows 4 and 5 (%)** | 1.081 | 0.2433 |
| **Preferred method?** | Yes | No |

**S6 Table. Simulation summary data (simulation conditions 2 of 11).** Comparison table of terrestrial organic microfossil concentration estimates ($c_M$) from Eqn 1 (when $M = L$) or Eqn 4 (when $M = F$), and their associated errors and sampling efforts from a simulated data set of randomly distributed target and exotic specimens.

Parameters: total targets in assemblage $= 30{,}000$; total markers in assemblage $= 25{,}000$; target-to-marker ratio = 6:5 (i.e., $\bar{\bar{u}} = 1.2$); $x$ count (linear method) $= 524$; simulated iterations $= 10^6$; $\omega = 2$; $N_{3C} = 17$; $N_{3F} = 20$; $\overline{Y}_3 = 27$; $\overline{Y}_3^* = 132$.

Since $\overline{Y}_3 < \overline{Y}_3^*$, the linear method is more efficient for this assemblage.

| Parameter estimates | Linear method ($M = L$) | FOVS method ($M = F$) |
|---|---|---|
| **1) Concentration ($c_M$; specimens/unit size), mean** | 30002 | 30000 |
| **2) Sampling effort ($e_M$; time units), mean** | 999.5 | 982.9 |
| **3) Estimated scaled standard error ($\widetilde{\sigma}_M$; %), mean (S16 and S17 Eqns)** | 6.540 | 6.611 |
| **4) Exact total standard error from true concentration ($\widetilde{\sigma}_{exact,M}$; %) (S19 and S20 Eqns)** | 6.555 | 6.560 |
| **5) Estimated scaled standard error, with finite population correction ($\widehat{\sigma}_M$; %), mean (S21 and S22 Eqns)** | 6.483 | 6.555 |
| **6) Difference between rows 4 and 5 (%)** | 1.091 | 0.07191 |
| **Preferred method?** | Yes | No |

**S7 Table. Simulation summary data (simulation conditions 3 of 11).** Comparison table of terrestrial organic microfossil concentration estimates ($c_M$) from Eqn 1 (when $M = L$) or Eqn 4 (when $M = F$), and their associated errors and sampling efforts from a simulated data set of randomly distributed target and exotic specimens.

Parameters: total targets in assemblage $= 30{,}000$; total markers in assemblage $= 20{,}000$; target-to-marker ratio = 3:2 (i.e., $\bar{\bar{u}} = 1.5$); $x$ count (linear method) $= 574$; simulated iterations $= 10^6$; $\omega = 2$; $N_{3C} = 17$; $N_{3F} = 25$; $\overline{Y}_3 = 27$; $\overline{Y}_3^* = 29.95$.

Since $\overline{Y}_3 < \overline{Y}_3^*$, the linear method is more efficient for this assemblage.

| Parameter estimates | Linear method ($M = L$) | FOVS method ($M = F$) |
|---|---|---|
| **1) Concentration ($c_M$; specimens/unit size), mean** | 30002 | 30000 |
| **2) Sampling effort ($e_M$; time units), mean** | 999.3 | 993.0 |
| **3) Estimated scaled standard error ($\widetilde{\sigma}_M$; %), mean (S16 and S17 Eqns)** | 6.628 | 6.646 |
| **4) Exact total standard error from true concentration ($\widetilde{\sigma}_{exact,M}$; %) (S19 and S20 Eqns)** | 6.643 | 6.586 |
| **5) Estimated scaled standard error, with finite population correction ($\widehat{\sigma}_M$; %), mean (S21 and S22 Eqns)** | 6.564 | 6.582 |
| **6) Difference between rows 4 and 5 (%)** | 1.185 | 0.05889 |
| **Preferred method?** | Yes | No |

**S8 Table. Simulation summary data (simulation conditions 4 of 11).** Comparison table of terrestrial organic microfossil concentration estimates ($c_M$) from Eqn 1 (when $M = L$) or Eqn 4 (when $M = F$), and their associated errors and sampling efforts from a simulated data set of randomly distributed target and exotic specimens.

Parameters: total targets in assemblage $= 30{,}000$; total markers in assemblage $= 15{,}000$; target-to-marker ratio = 2:1 (i.e., $\bar{\bar{u}} = 2$); $x$ count (linear method) $= 635$; simulated iterations $= 10^6$; $\omega = 2$; $N_{3C} = 17$; $N_{3F} = 33$; $\overline{Y}_3 = 27$; $\overline{Y}_3^* = 11.87$.

Since $\overline{Y}_3 > \overline{Y}_3^*$, the FOVS method is more efficient for this assemblage.

| Parameter estimates | Linear method ($M = L$) | FOVS method ($M = F$) |
|---|---|---|
| **1) Concentration ($c_M$; specimens/unit size), mean** | 30004 | 30000 |
| **2) Sampling effort ($e_M$; time units), mean** | 999.6 | 1005 |
| **3) Estimated scaled standard error ($\widetilde{\sigma}_M$; %), mean (S16 and S17 Eqns)** | 6.866 | 6.699 |
| **4) Exact total standard error from true concentration ($\widetilde{\sigma}_{exact,M}$; %) (S19 and S20 Eqns)** | 6.887 | 6.630 |
| **5) Estimated scaled standard error, with finite population correction ($\widehat{\sigma}_M$; %), mean (S21 and S22 Eqns)** | 6.793 | 6.623 |
| **6) Difference between rows 4 and 5 (%)** | 1.368 | 0.09921 |
| **Preferred method?** | No | Yes |

**S9 Table. Simulation summary data (simulation conditions 5 of 11).** Comparison table of terrestrial organic microfossil concentration estimates ($c_M$) from Eqn 1 (when $M = L$) or Eqn 4 (when $M = F$), and their associated errors and sampling efforts from a simulated data set of randomly distributed target and exotic specimens.

Parameters: total targets in assemblage $=$ 30,000; total markers in assemblage $=$ 10,000; target-to-marker ratio = 3:1 (i.e., $\bar{\bar{u}} = 3$); $x$ count (linear method) $=$ 711; simulated iterations $= 10^6$; $\omega = 2$; $N_{3C} = 17$; $N_{3F} = 47$; $\overline{Y}_3 = 27$; $\overline{Y}_3^* = 5.687$.

Since $\overline{Y}_3 > \overline{Y}_3^*$, the FOVS method is more efficient for this assemblage.

| **Parameter estimates** | **Linear method** ($M = L$) | **FOVS method** ($M = F$) |
|---|---|---|
| **1) Concentration ($c_M$; specimens/unit size), mean** | 30000 | 30000 |
| **2) Sampling effort ($e_M$; time units), mean** | 1001 | 1010 |
| **3) Estimated scaled standard error ($\widetilde{\sigma}_M$; %), mean (S16 and S17 Eqns)** | 7.479 | 6.804 |
| **4) Exact total standard error from true concentration ($\widetilde{\sigma}_{exact,M}$; %) (S19 and S20 Eqns)** | 7.508 | 6.709 |
| **5) Estimated scaled standard error, with finite population correction ($\widehat{\sigma}_M$; %), mean (S21 and S22 Eqns)** | 7.390 | 6.703 |
| **6) Difference between rows 4 and 5 (%)** | 1.573 | 0.09027 |
| **Preferred method?** | No | Yes |

**S10 Table. Simulation summary data (simulation conditions 6 of 11).** Comparison table of terrestrial organic microfossil concentration estimates ($c_M$) from Eqn 1 (when $M = L$) or Eqn 4 (when $M = F$), and their associated errors and sampling efforts from a simulated data set of randomly distributed target and exotic specimens.

Parameters: total targets in assemblage $=$ 30,000; total markers in assemblage $=$ 5,000; target-to-marker ratio = 6:1 (i.e., $\bar{\bar{u}} = 6$); $x$ count (linear method) $=$ 806; simulated iterations $= 10^6$; $\omega = 2$; $N_{3C} = 16$; $N_{3F} = 83$; $\overline{Y}_3 = 27$; $\overline{Y}_3^* = 2.693$.

Since $\overline{Y}_3 > \overline{Y}_3^*$, the FOVS method is more efficient for this assemblage.

| Parameter estimates | Linear method ($M = L$) | FOVS method ($M = F$) |
|---|---|---|
| **1) Concentration ($c_M$; specimens/unit size), mean** | 30004 | 29995 |
| **2) Sampling effort ($e_M$; time units), mean** | 1000 | 1003 |
| **3) Estimated scaled standard error ($\tilde{\sigma}_M$; %), mean (S16 and S17 Eqns)** | 9.331 | 7.114 |
| **4) Exact total standard error from true concentration ($\tilde{\sigma}_{exact,M}$; %) (S19 and S20 Eqns)** | 9.454 | 6.949 |
| **5) Estimated scaled standard error, with finite population correction ($\hat{\sigma}_M$; %), mean (S21 and S22 Eqns)** | 9.206 | 6.946 |
| **6) Difference between rows 4 and 5 (%)** | 2.632 | 0.03893 |
| **Preferred method?** | No | Yes |

**S11 Table. Simulation summary data (simulation conditions 7 of 11).** Comparison table of terrestrial organic microfossil concentration estimates ($c_M$) from Eqn 1 (when $M = L$) or Eqn 4 (when $M = F$), and their associated errors and sampling efforts from a simulated data set of randomly distributed target and exotic specimens.

Parameters: total targets in assemblage $= 30{,}000$; total markers in assemblage $= 3{,}000$; target-to-marker ratio = 10:1 (i.e., $\bar{\bar{u}} = 10$); $x$ count (linear method) $= 852$; simulated iterations $= 10^6$; $\omega = 2$; $N_{3C} = 15$; $N_{3F} = 119$; $\overline{Y}_3 = 27$; $\overline{Y}_3^* = 1.803$.

Since $\overline{Y}_3 > \overline{Y}_3^*$, the FOVS method is more efficient for this assemblage.

| Parameter estimates | Linear method ($M = L$) | FOVS method ($M = F$) |
|---|---|---|
| **1) Concentration ($c_M$; specimens/unit size), mean** | 29997 | 30000 |
| **2) Sampling effort ($e_M$; time units), mean** | 1000 | 994.3 |
| **3) Estimated scaled standard error ($\tilde{\sigma}_M$; %), mean (S16 and S17 Eqns)** | 11.49 | 7.490 |
| **4) Exact total standard error from true concentration ($\tilde{\sigma}_{exact,M}$; %) (S19 and S20 Eqns)** | 11.74 | 7.251 |
| **5) Estimated scaled standard error, with finite population correction ($\hat{\sigma}_M$; %), mean (S21 and S22 Eqns)** | 11.29 | 7.241 |
| **6) Difference between rows 4 and 5 (%)** | 3.885 | 0.1453 |
| **Preferred method?** | No | Yes |

**S12 Table. Simulation summary data (simulation conditions 8 of 11).** Comparison table of terrestrial organic microfossil concentration estimates ($c_M$) from Eqn 1 (when $M = L$) or Eqn 4 (when $M = F$), and their associated errors and sampling efforts from a simulated data set of randomly distributed target and exotic specimens.

Parameters: total targets in assemblage $= 30{,}000$; total markers in assemblage $= 2{,}000$; target-to-marker ratio = 15:1 (i.e., $\bar{\bar{u}} = 15$); $x$ count (linear method) $= 877$; simulated iterations $= 10^6$; $\omega = 2$; $N_{3C} = 14$; $N_{3F} = 154$; $\overline{Y}_3 = 27$; $\overline{Y}_3^* = 1.363$.

Since $\overline{Y}_3 > \overline{Y}_3^*$, the FOVS method is more efficient for this assemblage.

| **Parameter estimates** | **Linear method** ($M = L$) | **FOVS method** ($M = F$) |
|---|---|---|
| **1) Concentration ($c_M$; specimens/unit size), mean** | 30004 | 29999 |
| **2) Sampling effort ($e_M$; time units), mean** | 1001 | 991.2 |
| **3) Estimated scaled standard error ($\tilde{\sigma}_M$; %), mean (S16 and S17 Eqns)** | 13.66 | 7.918 |
| **4) Exact total standard error from true concentration ($\tilde{\sigma}_{exact,M}$; %) (S19 and S20 Eqns)** | 14.26 | 7.593 |
| **5) Estimated scaled standard error, with finite population correction ($\hat{\sigma}_M$; %), mean (S21 and S22 Eqns)** | 13.46 | 7.574 |
| **6) Difference between rows 4 and 5 (%)** | 5.634 | 0.2542 |
| **Preferred method?** | No | Yes |

**S13 Table. Simulation summary data (simulation conditions 9 of 11).** Comparison table of terrestrial organic microfossil concentration estimates ($c_M$) from Eqn 1 (when $M = L$) or Eqn 4 (when $M = F$), and their associated errors and sampling efforts from a simulated data set of randomly distributed target and exotic specimens.

Parameters: total targets in assemblage $= 30{,}000$; total markers in assemblage $= 1{,}500$; target-to-marker ratio = 20:1 (i.e., $\bar{\bar{u}} = 20$); $x$ count (linear method) $= 890$; simulated iterations $= 10^6$; $\omega = 2$; $N_{3C} = 14$; $N_{3F} = 180$; $\overline{Y}_3 = 27$; $\overline{Y}_3^* = 1.132$.

Since $\overline{Y}_3 > \overline{Y}_3^*$, the FOVS method is more efficient for this assemblage.

| Parameter estimates | Linear method $(M = L)$ | FOVS method $(M = F)$ |
|---|---|---|
| **1) Concentration ($c_M$; specimens/unit size), mean** | 30004 | 30000 |
| **2) Sampling effort ($e_M$; time units), mean** | 1001 | 1009 |
| **3) Estimated scaled standard error ($\widetilde{\sigma}_M$; %), mean (S16 and S17 Eqns)** | 15.43 | 8.306 |
| **4) Exact total standard error from true concentration ($\widetilde{\sigma}_{exact,M}$; %) (S19 and S20 Eqns)** | 16.40 | 7.890 |
| **5) Estimated scaled standard error, with finite population correction ($\widehat{\sigma}_M$; %), mean (S21 and S22 Eqns)** | 15.20 | 7.866 |
| **6) Difference between rows 4 and 5 (%)** | 7.310 | 0.2988 |
| **Preferred method?** | No | Yes |

**S14 Table. Simulation summary data (simulation conditions 10 of 11).** Comparison table of terrestrial organic microfossil concentration estimates ($c_M$) from Eqn 1 (when $M = L$) or Eqn 4 (when $M = F$), and their associated errors and sampling efforts from a simulated data set of randomly distributed target and exotic specimens.

Parameters: total targets in assemblage $= 30{,}000$; total markers in assemblage $= 1{,}000$; target-to-marker ratio = 30:1 (i.e., $\bar{\bar{u}} = 30$); $x$ count (linear method) $= 903$; simulated iterations $= 10^6$; $\omega = 2$; $N_{3C} = 13$; $N_{3F} = 219$; $\overline{Y}_3 = 27$; $\overline{Y}_3^* = 0.8821$.

Since $\overline{Y}_3 > \overline{Y}_3^*$, the FOVS method is more efficient for this assemblage.

| Parameter estimates | Linear method $(M = L)$ | FOVS method $(M = F)$ |
|---|---|---|
| **1) Concentration ($c_M$; specimens/unit size), mean** | 29999 | 30000 |
| **2) Sampling effort ($e_M$; time units), mean** | 1000 | 1012 |
| **3) Estimated scaled standard error ($\widetilde{\sigma}_M$; %), mean (S16 and S17 Eqns)** | 18.65 | 9.011 |
| **4) Exact total standard error from true concentration ($\widetilde{\sigma}_{exact,M}$; %) (S19 and S20 Eqns)** | 20.57 | 8.441 |
| **5) Estimated scaled standard error, with finite population correction ($\widehat{\sigma}_M$; %), mean (S21 and S22 Eqns)** | 18.37 | 8.407 |
| **6) Difference between rows 4 and 5 (%)** | 10.67 | 0.4088 |
| **Preferred method?** | No | Yes |

**S15 Table. Simulation summary data (simulation conditions 11 of 11).** Comparison table of terrestrial organic microfossil concentration estimates ($c_M$) from Eqn 1 (when $M = L$) or Eqn 4 (when $M = F$), and their associated errors and sampling efforts from a simulated data set of randomly distributed target and exotic specimens.

Parameters: total targets in assemblage $= 30{,}000$; total markers in assemblage $= 500$; target-to-marker ratio = 60:1 (i.e., $\bar{\bar{u}} = 60$); $x$ count (linear method) $= 917$; simulated iterations $= 10^6$; $\omega = 2$; $N_{3C} = 11$; $N_{3F} = 283$; $\overline{Y}_3 = 27$; $\overline{Y}_3^* = 0.5888$.

Since $\overline{Y}_3 > \overline{Y}_3^*$, the FOVS method is more efficient for this assemblage.

| Parameter estimates | Linear method ($M = L$) | FOVS method ($M = F$) |
|---|---|---|
| **1) Concentration ($c_M$; specimens/unit size), mean** | 30001 | 30001 |
| **2) Sampling effort ($e_M$; time units), mean** | 1000 | 1012 |
| **3) Estimated scaled standard error ($\widetilde{\sigma}_M$; %), mean (S16 and S17 Eqns)** | 26.32 | 10.73 |
| **4) Exact total standard error from true concentration ($\widetilde{\sigma}_{exact,M}$; %) (S19 and S20 Eqns)** | 33.33 | 9.809 |
| **5) Estimated scaled standard error, with finite population correction ($\widehat{\sigma}_M$; %), mean (S21 and S22 Eqns)** | 25.93 | 9.721 |
| **6) Difference between rows 4 and 5 (%)** | 22.22 | 0.8973 |
| **Preferred method?** | No | Yes |

# Supporting information 16

## Matlab code for simulations

The codes used to generate the simulation data in this paper have not been optimised, and have some components that are either not used or not fully implemented. However, in the interests of full transparency, we include the exact versions of the code that we used for our results below.

### Data for S5–S15 Tables and S17 Fig

Code:

- **[Main] BigFossilSimsV3.m**
    - **[Dependent] MicrofossilSimV3.m**
    - **[Dependent] MicrofossilSim_iV3.m**
    - **[Dependent] FOVoptimiserV1.m**

Use:

Specify the following variables in **BigFossilSimsV3.m**:

- [Line 3] **its**: the number of independent Monte Carlo instances to generate for each set of parameters.
- [Line 22] **params**: **[Mx, Mn, tab]**
    - **Mx**: the total number of targets on each virtual study area.
    - **Mn**: the total number of markers on each virtual study area.
    - **tab**: value of the dose error used in Eqns 2 and 5.

- Note: Multiple rows of this variable can be specified to run multiple batches, via: **[(first batch parameters); (second batch parameters); ...]**

- [Line 27] **alpha**: this is the field of view transition factor ($\omega$). Default is $\omega = 2$.
- [Line 33] **work**: the fixed value of work that the program tries to achieve for each method.
    - Linear method: Eqn 9 is used to choose the number of targets to count.
    - FOVS method: Eqns 14 and 15 are used to choose the optimal number of calibration and full count fields of view, via the code **FOVoptimiserV1.m**.

Notes:

- Ensure that there is a **\SimData** subdirectory for the program in which to store the data files.
- The command line output will be saved in a file called **BigFossilSimsV3_Opt_TX_itsY.txt**, where
    - **X** is 10000 times the **tablet error** (to ensure an integer); and
    - **Y** is the value of **its**.

## Data for Fig 3

Code:

- **[Main] SimStatsChecker.m**
    - **[Dependent] MicrofossilSim_iCheck.m**

Use:

Call the function **SimStatsChecker(Mx,Mn,tlim,fn,its,fopt)**, where the arguments are:

- **Mx**: The total number of targets on each virtual slide.
- **Mn**: The total number of markers on each virtual slide.
- **tlim**: (Linear method) the number of targets to count in the window.
- **fn**: (FOVS method) the number of full count fields of view in which to count markers.
- **its**: The number of independent Monte Carlo instances to generate.
- **fopt**: Not used. Set to 1.

## Data for Fig 8

Code:

- **[Main] PrecWRTWorkV2.m**
  - **[Dependent] WorkSimV2.m**
  - **[Dependent] WorkSimV2_i.m**

Use:

Specify the following variables in **PrecWRTWorkV2.m**

- [Line 7] **its**: The number of independent Monte Carlo instances to generate for each set of parameters.
- [Line 8] **bigfx**: The number of calibration counts for the "high calibration counts" sequence in Fig 8 (black plus).
- [Line 9] **medfx**: The number of calibration counts for the "medium calibration counts" sequence in Fig 8 (blue stars).

- [Line 10] **smallfx**: The number of calibration counts for the "low calibration counts" sequence in Fig 8 (red stars).
- [Lines 19–24] **params**: **[Mx, Mx, tlim, fnmax, omega]**
  - **Mx**: The total number of targets on each virtual slide.
  - **Mn**: The total number of markers on each virtual slide.
  - **tlim**: (Linear method) the number of targets to count in the window.
  - **fnmax**: Not used. Set to 1.
  - **omega**: This is the field of view transition factor ($\omega$).
  - Note: Multiple rows of this variable can be specified to run multiple batches, via: **[(first batch parameters); (second batch parameters); ...]**

The simulations currently assume that the marker dose (e.g., tablet of *Lycopodium* spores) error is zero, i.e.: $\left(\frac{s_{1P}}{\sqrt{N_1}}\right)^2 = 0$. If you wish to increase this, then change the following variable:

- [**WorkSimV2_i.m**, Line 33] **tab**: Value of the marker dose error used in Eqns 2 and 5.

Notes:

- Ensure that there is a **\SimData** subdirectory for the program to store the data files in.
- The command line output will be saved in a file called **WorkSimOpt_tab0_itsY.txt**, where:
  - **Y** is the value of **its**.

  **tab0** records that the marker dose error is zero for the simulations. This is hard-coded and will not update if the value of tab is changed in **WorkSimV2_i.m**.